\documentclass[twocolumn,pra,superscriptaddress]{revtex4}

\usepackage{graphicx}
\usepackage{amsmath,amssymb}
\usepackage{hyperref}
\usepackage[latin1]{inputenc}

\newcommand{\figwidth}{0.95\columnwidth}
\newcommand{\largefigwidth}{0.85\columnwidth}
\newcommand{\Tr}{\operatorname{Tr}}

\newcommand{\pd}{\phantom{\dagger}}
\newcommand{\sign}{\operatorname{sign}}
\renewcommand{\Re}{\operatorname{Re}}

\begin{document}
\title{Quantum quench dynamics of the Luttinger model}
\author{A.~Iucci}
\affiliation{Instituto de Física la Plata (IFLP) - CONICET and Departamento de Física,\\
Universidad Nacional de La Plata, CC 67, 1900 La Plata, Argentina}
\affiliation{DPMC-MaNEP, University of Geneva, 24 Quai Ernest Ansermet CH-1211 Geneva 4, Switzerland}
\affiliation{Donostia International Physics Center (DIPC), Manuel de Lardiz\'abal 4, 20018 San Sebasti\'an, Spain}
\author{M.~A.~Cazalilla}
\affiliation{Centro de Física de Materiales (CFM). Centro Mixto CSIC-UPV/EHU. Edificio Korta, Avenida de Tolosa
72, 20018 San Sebasti\'an,  Spain}
\affiliation{Donostia International Physics Center (DIPC), Manuel de Lardiz\'abal 4, 20018 San Sebasti\'an, Spain}
\begin{abstract}
The dynamics of the Luttinger model after a quantum quench is studied. We compute in detail one and two-point correlation functions for two types of quenches: from a non-interacting to an interacting Luttinger model and vice-versa.  In the former case,  the non-interacting Fermi gas features in the momentum distribution and other correlation functions are destroyed  as time evolves.  In the infinite-time limit,  equal-time correlations are power-laws but  the critical exponents are found to differ from their equilibrium values. In all cases, we find that these correlations  are well described by a generalized Gibbs ensemble [M. Rigol \emph{et al.} Phys. Rev. Lett. \textbf{98}, 050405 (2007)], which assigns a momentum dependent temperature to each eigenmode.
\end{abstract}
\maketitle
\section{Introduction}

Most of the theoretical effort in the field of strongly correlated
quantum systems over the past few decades has focused on
understanding the equilibrium properties of these fascinating
systems. For instance, achieving a complete understanding of the
phase diagram of rather ``simple'' models like the
two-dimensional fermionic Hubbard model still remains a huge
challenge. Nevertheless, however important these endeavors are,
understanding the phase diagram and the equilibrium properties of the phases
of strongly correlated systems will not certainly exhaust the possibilities for finding new and
surprising phenomena in these complex systems, especially out of
equilibrium.

In classical systems, the existence of steady  states out of equilibrium
is well known. Very often, however, the properties of such  states
have very little to do with the equilibrium properties of the systems
where they occur. Moreover, also very often  their
existence cannot be inferred from
any previous knowledge about the  equilibrium phase diagram: They
are \emph{emergent} phenomena.

 One good example of a classical non-equilibrium steady state
 is provided by the appearance of  Rayleigh-B{\'e}nard  convection cells  when a
fluid layer is driven out of equilibrium by  a
temperature gradient. Indeed, it is known that
dissipation plays an important role in the formation of these classical
non-equilibrium  states. However, different from
classical systems, dissipation in
quantum systems causes decoherence, which  usually destroys
any interesting quantum interference effects. Thus, although one may also wonder if
non-equilibrium steady states can also appear
when quantum systems are driven out of equilibrium,
the study of non-equilibrium phenomena in quantum many-body systems has been  regarded, until
very recently, as a subject of mostly academic interest. The reason for this
may be  decoherence due to  coupling with the environment, which is always present in
most experimental realizations of large quantum many-body systems,
and which prevents the observation of \emph{coherent} quantum
evolution for long times.

 However, the recent availability  of highly controlable systems of
 ultracold atomic gases has finally provided the largely lacking
experimental motivation for the study of non-equilibrium
phenomena, leading to an explosion of theoretical
activity.\cite{altman_quench_2002,sengupta_quench_QCP,barankov_%
dynamical_projection,yuzbashyan_BCS_quench,kollath_density_waves,%
altman_projection_feshbach_molecules,ruschhaupt_quench_momentum_%
interference,yuzbashyan_quench_fermions_order_parameter,cazalilla_%
quench_LL,rigol_generalized_gibbs_hcbosons,perfetto_quench_coupled_LL, %
calabrese_quench_CFT,Rigol_hc_noneq2,manmana_quench_spinless_fermions_nnn,
calabrese_quench_CFT_long,kollath_quench_BH,Gritsev_spectroscopy_07,%
eckstein_generalized_gibbs_FK,eckstein_generalized_gibbs_hubbard,%
kehrein_quench_hubbard,eisert_crap1,eisert_crap2,rigol_noninteg,Manmana_long,%
degrandi_adiabatic_quench,Reimann08,faribault_fermion_pairing_model,Patane08,Patane08b,%
Rossini08,Barmettler08,sen_nonlinear_quench_QCP,sen_2,sen_3,sen_4,sen_5,sen_6,sen_7,roux_time_dependent_lanczos} Ultracold atomic gases are especially
interesting because they are very weakly coupled to the
environment, thus remaining fully quantum coherent for fairly long
times (compared to the typical duration of an experiment). At the
same time, it is relatively easy to measure the \emph{coherent}
evolution in time of observables such as the density or the
momentum distribution. Thus,  theorists can  now begin to
pose questions such as: Assuming that a many-body system is prepared in a
given initial state that is not an eigenstate of  the Hamiltonian,
how will it evolve in time? And, more specifically,
will it reach a stationary or quasi-stationary state? If so,
what will be the properties of such a state?
How much memory will the system retain of its initial conditions?

From another point of view, the problem described in the previous
paragraph can be formulated as the study of the response of a
system to a sudden perturbation in which the Hamiltonian is
changed over a time scale much
shorter than any other characteristic time scales of the system.
In what follows, we shall refer to this type  of experiment as a
\emph{quantum quench}. Quantum quenches are also of particular
interest to the `quantum engineering' program for ultracold atomic
gases.\cite{quantum_engineering}
The reason is that, if we intend to use these
highly tunable and controllable systems as
quantum simulators of models  of many-body
physics [such as the two-dimensional (2D) fermionic Hubbard model mentioned above],
it is  utterly important to understand  to what extent
the final state of the quantum simulator depends on the  state in
which it was initially prepared.  In particular, one is interested
in finding out whether the observables in the final state state
can be obtained from a standard statistical ensemble (say, the
microcanonical, or the canonical ensemble at an effective
temperature). If this is so, one would speak of
\emph{thermalization}. If this does not happen, then how much
memory does  the system retain about its initial state beyond the
average energy  $E =  \langle H \rangle$?

We would like to
emphasize that the above questions are not a merely academic.
Indeed, ultracold atomic systems allow for the study of non adiabatic dynamics
when the system is driven between two quantum phases such as
a superfluid and a Mott insulator.
\cite{greiner_transition_superfluid_mott,greiner_fast_tunnability}
Also, in a recent experiment, \cite{kinoshita_non_thermalization}
it has been shown that a  faithful realization of the Lieb-Liniger
model \cite{lieb_liniger_model} exhibits absence of
thermalization. In other words, when prepared in an
non-equilibrium state, the experimental system reached a steady
state that cannot be described by any  of the `standard' ensembles of
Statistical Mechanics. This absence of thermalization seems to be
a consequence of the \emph{integrability} of the Lieb-Liniger
model, that is, the existence of an infinite number of
independently conserved quantities.  This conclusion
was backed by the theoretical analysis of Rigol and
coworkers,\cite{rigol_generalized_gibbs_hcbosons} who
noticed that the non-equilibrium dynamics of an integrable
system is highly constrained. Thus, based
on numerical simulations for the Tonks-Girardeau limit
of the Lieb-Liniger model, these authors conjectured
that the long-time values of  some observables should converge to those
obtained from a generalized Gibbs ensemble, which
can be constructed using the maximum (von Neuman-) entropy
principle.~\cite{JaynesI, JaynesII,rigol_generalized_gibbs_hcbosons}
The conjecture was first analytically confirmed  by analyzing
an interaction quench in the Luttinger
model by one of us.~\cite{cazalilla_quench_LL}
Later, it has been also found true  in other
integrable models: Cardy and Calabresse studied
a quench in a Harmonic chain,~\cite{calabrese_quench_CFT_long}
Eckstein and Kollar analyzed
the Falikov-Kimball model in
infinite dimensions,\cite{eckstein_generalized_gibbs_FK} and
the $1/r$  Hubbard model in one dimension.~\cite{eckstein_generalized_gibbs_hubbard}
Moeckel and Kehrein~\cite{kehrein_quench_hubbard}
studied an interaction quench in the Hubbard model
in infinite dimensions  by a flow equation method,
and found that the system reaches
an intermediate non-thermal state. Finally, recent numerical
studies also have suggested  that lack of thermalization may even
persist in the absence of integrability in one-dimensional
systems,\cite{manmana_quench_spinless_fermions_nnn}
or that  it  may occur only certain parameter regimes of
non-integrable models.\cite{kollath_quench_BH}

However, it can be expected  that,~\cite{Reimann08}
for a rather general choice of the
initial state, along with a situation where there are few
conserved quantities, the system will lose memory of most
of the details of the initial state and, after it reaches a
steady state, the expectation of
many experimentally accessible observables such as
the particle density or the momentum distribution, will
look essentially  identical to those obtained from a standard
thermal ensemble.\footnote{Implicit in this
discussion it is the fact we are attempting at a description of the system as a
whole, not separating its degrees of freedom into a `system'  and
a `reservoir'. We believe that this point of view is more
appropriate when discussing ultracold atomic systems, given that they
are very weakly coupled to the environment.}
Indeed, this is what seems to
be observed  in the vast majority of the experiments with
ultracold atomic gases. However, for experimental many-body
systems in general, it is hard to quantify whether this will
be always the case. We should take into account
that (except perhaps in the case of ultracold atomic gases)
the exact form of the quantum Hamiltonian is frequently
not known with accuracy. And even when it is known, it is not always
possible to tell \emph{a priori} whether the system
is integrable or even if it has other conserved quantities besides
the ones assumed by the standard thermodynamic
ensembles.  As a possible experimental check,
we can say that,  provided the final result is largely
independent of the  particular
details of the preparation of the initial state and that the observations
agree with those obtained from a thermal ensemble,
we can say that thermalization has occurred.
Indeed,  some recent numerical evidence,\cite{rigol_noninteg}
supplemented by the extension to many-particle systems
of a conjecture known as `eigenstate thermalization hypothesis'
(first introduced in the context of quantum chaos\cite{Srednicki_eigenstate}),
seems to indicate that  lack of integrability will in general lead
to thermalization (in the sense defined above). Indeed, Reimann~\cite{Reimann08} has
recently analytically demonstrated that, under realistic experimental conditions,
equilibration will be observed in an isolated system that has been initially
prepared in an non-equilibrium mixed state.
Nevertheless, even if the issue of thermalization for non-integrable systems
may have been settled, other questions such as the details of the
transition from the integrable case (which thermalizes to a
generalized Gibbs ensemble) to the non integrable
case, which thermalizes to the standard microcanonical
or Gibbs ensemble for large enough systems, are questions
that are still far from being completely understood.~\footnote{This question
is also related to the problems concerning
the applicability of the maximum entropy approach~\cite{JaynesI,JaynesII}
to Statistical Mechanics. See for instance the critique
by Ma in Ref.~\onlinecite{ma_stat_mech}. However, the
maximum entropy approach is advocated by Balian.~\cite{balian_stat_mech}}

In this article, we will not try to answer the difficult questions
posed in the previous paragraph. Instead, we focus on analyzing
the quench dynamics of a relatively well-known
one dimensional model, namely, the Luttinger model (LM).  A brief
account of the results in this article  has  been already
published elsewhere.\cite{cazalilla_quench_LL}
In a future publication, we shall also deal with another closely
related model, the sine-Gordon model.~\cite{unpub}

 The LM Hamiltonian can be represented as a quadratic form of creation and destruction
operators.  The  long time behavior following a quench for Hamiltonians of this form
has been recently considered
by Barthel and Scholw\"ock.\cite{schollwoeck_quench}
These authors provided some general
conditions for the appearance of dephasing and
steady non-thermal states. This question has been
also taken up recently by Kollar and
Eckstein.~\cite{eckstein_generalized_gibbs_hubbard}
However, since the Luttinger model
may be relevant   to experiments using ultracold atomic gases
(see Sect.~\ref{sec:exp}) or numerical simulations,
it is important to obtain analytical results.
The simplicity of this model also allows us
to test in detail a number of  general
results.~\cite{calabrese_quench_CFT,schollwoeck_quench}

The rest of this article is organized as follows:
In Sect.~\ref{sec:LM}, we discuss the evolution of some simple correlation
functions of the Luttinger model. We consider the case
where the interactions between the fermions is suddenly switched
on, and the reverse situation, that is, when the interaction in suddenly switched off.
In Sect.~\ref{sec:generalized}, we discuss how the infinite-time behavior
of some of the correlation functions following a quantum quench can be
obtained from a generalized Gibbs ensemble. We also discuss some
observables for which this conjecture fails.
The experimental relevance of our results is
briefly discussed in Sect.~\ref{sec:exp}, along with
other conclusions of this work. Finally, the details of
some of the lengthier calculations are provided
in Appendixes A to C.

\section{The Luttinger model}\label{sec:LM}

The Luttinger model (LM) describes a one-dimensional (1D) system of interacting fermions
with linear dispersion. It was first introduced by Luttinger \cite{luttinger_model}
but its complete solution was only later obtained
by Mattis and Lieb,\cite{mattis_lieb_luttinger_model}
who showed that the elementary excitations of the system are not
fermionic quasi-particles. Instead, Mattis and Lieb introduced a set of bosonic fields
describing collective density modes (phonons) of the system, which are the true
elementary low-energy excitations  of the LM. The methods of Mattis and
Lieb bear strong resemblance to the early work of Tomonaga~\cite{Tomonaga_1D_electron_gas}
on the one-dimensional electron gas. Extending the work of Tomonaga, as weel as that of Mattis and Lieb,
Luther and Peschel \cite{luther_peschel_correlation_functions}
computed the  one and two-particle correlation functions in equilibrium,
thus showing that correlations exhibit (at zero temperature and long distances)
a non-universal power-law behavior signaling the absence
of long-range order.  Later, Haldane \cite{haldane_exponents_spin_chain,%
haldane_luttinger_liquid,haldane_effective_harmonic_fluid_approach}
conjectured that these properties (\emph{i.e.} collective elementary excitations exhausting
the low-energy part of the spectrum as well as power-law correlations) are
distinctive features of a large class of gapless interacting one-dimensional systems that
he termed (Tomonaga-)`Luttinger liquids'. Using  the modern language of critical phenomena,
the LM can be understood
as a  fixed point of the renormalization-group for a large class of gapless many-body
systems in one dimension: the equilibrium properties at low temperatures of many 1D systems
are \emph{universal} in the sense that they can be accurately described by the LM.
However, in this work we shall be concerned
with  non-equilibrium properties of the LM,  and  because the latter
can involve highly excited states,  we shall make no claim for universality.
The precise conditions  under which the results obtained here apply to
real systems that are in the Tomonaga-Luttinger class should be investigated carefully
in each particular  instance (see discussion in Sect.~\ref{sec:exp}).

The Hamiltonian of the LM can be written as follows:
\begin{align}
H_\text{LM} &=  H_0 + H_2 + H_4,  \label{eq:hlm1}\\
H_0 &= \sum_{p,\alpha = r,l} \hbar v_F p :\psi^{\dag}_{\alpha}(p)
\psi_{\alpha}(p):\,, \label{eq:hlm2}\\
H_2 &=  \frac{2\pi \hbar}{L}\sum_{q} g_2(q) :J_r(q) J_l(q):\,, \label{eq:hlm3} \\
H_4 &= \frac{\pi\hbar}{L} \sum_{q, \alpha = r,l} g_4(q)
\, :J_{\alpha}(q) J_{\alpha}(-q): \,.\label{eq:hlm4}
\end{align}
Here $\psi^{\phantom{\dagger}}_{\alpha}(p)$ and $\psi^{\dagger}_{\alpha}(p)$
are fermion creation and annihilation operators at some momentum $p$ and
$J_{\alpha}(q) = \sum_{p} :\psi^{\dag}_{\alpha}(p+q) \psi_{\alpha}(p):$.
The index $\alpha = r,l$ refers to the \emph{chirality} of the
fermion species, which can be either right ($r$) or left ($l$)
moving; the symbol $:\ldots:$ stands for normal ordering prescription
for fermionic operators. This is needed to remove from the
expectation values the infinite contributions arising
from the fact that the ground state is a Dirac sea,\cite{haldane_luttinger_liquid}
namely a state where all single-particle fermion levels with $p < 0$
are occupied for both chiralities. This defines a stable ground state
(at the non-interacting level), which in what follows
will be denoted by $| 0\rangle$.

\subsection{Bosonization solution of the LM}

 In this section we briefly review the solution of the
LM. The Hamiltonian in Eqs.~(\ref{eq:hlm1}) to (\ref{eq:hlm4})
can be written as a quadratic Hamiltonian in
terms of a set of bosonic operators.\cite{mattis_lieb_luttinger_model}
First we note that the density operators $J_{\alpha}(q) = \sum_{p}
:\psi^{\dag}_{\alpha}(p+q) \psi_{\alpha}(p):$ obey the following
commutation rules:
\begin{align}
\left[J_{\alpha}(q),J_{\beta}(q') \right] = \left(\frac{qL}{2\pi}
\right) \delta_{q+q',0} \delta_{\alpha\beta},
\end{align}
which can be transformed into the Heisenberg algebra of the familiar
bosonic operators by introducing:
\begin{equation}
b(q) = - i \left(\frac{2\pi}{|q|L}\right)^{1/2} \left[ \vartheta(q)
J_r(-q) - \vartheta(-q) J_l(q) \right].\label{eq:bosons}
\end{equation}
where $\vartheta(q)$ is the step function. Note that the $q = 0$
components (known as  `zero modes') require a separate treatment since
$J_{\alpha}(0) = N_{\alpha}$ is the deviation, relative to the ground state,
in the number of fermions of chirality $\alpha = r, l$.  However, rather than working
with $N_r$ and $N_l$, it is convenient to introduce:
\begin{equation}
N = N_r + N_l \quad J = N_r - N_l,
\end{equation}
which, since $N_r$ and $N_l$ are integers, must obey the following
selection rule $(-1)^{N} = (-1)^J$ when the Fermi fields obey anti-periodic boundary
conditions: $\psi_{\alpha}(x + L) = -\psi_{\alpha}(x)$ ($L$ is the length of the system). Therefore  $\psi_{\alpha}(x) = L^{-1/2}\sum_{p} e^{- a_0 |p|} e^{i s_{\alpha} p
x} \psi_{\alpha}(p)$, being $s_{r} = - s_{l} = +1$ and  $p = 2(n-\frac{1}{2})\pi/L$, where $n$ is an
integer, and  $a_0 \to 0^{+}$.\cite{haldane_luttinger_liquid}

The Hamiltonian $H_\text{LM}$ can be expressed in terms of the bosonic
operators introduced in Eq.~(\ref{eq:bosons}):
\begin{align}
H_0 &= \sum_{q \neq 0} \hbar v_F |q| \,  b^\dag(q) b(q)
+\frac{\hbar \pi v_F}{2L} \left( N^2 + J^2 \right), \\
H_2 &= \frac{1}{2} \sum_{q\neq 0} g_2(q)|q| \, \left[ b(q)
b(-q)
+ b^\dag(q) b^\dag(-q) \right]\notag\\
&\qquad\qquad+\frac{\hbar \pi g_2(0)}{2L} \left( N^2 - J^2 \right),\\
H_4 &= \sum_{q\neq 0} \hbar g_4(q) |q|\,  b^\dag(q) b(q) +
\frac{\hbar \pi g_4(0)}{2L} (N^2 + J^2).
\end{align}

Ignoring the zero mode part, the above Hamiltonian has
the form of Eq.~(\ref{eq:genham}), with the following identifications: $
\omega_0(q)= v_F |q|$, $m(q,t)  =  g_4(q)|q|$, and $g(q,t)=
g_2(q)|q|$, and it can be therefore be brought into diagonal
form by means of the canonical transformation of Eq.~(\ref{eq:bogol}).
Hence, the Hamiltonian  takes the form of Eq.~(\ref{eq:ham_diag}) with
$\omega(q)=v(q)|q|$, being
$v(q)=\{[v_F+g_4(q)]^2-[g_2(q)]^2\}^{1/2}$, and $q \neq 0$. As to the
zero mode contribution:
\begin{equation}
H_\text{ZM}=\frac{\hbar\pi v_N}{2L}N^2+\frac{\hbar\pi v_J}{2L}J^2,
\end{equation}
where $v_N=v_F+g_4(0)+g_2(0)$ and $v_J=v_F+g_4(0)-g_2(0)$.
This defines the equilibrium solution of the LM. In the following
sections we shall be concerned with the quench dynamics of this model.

\subsection{Suddenly turning-on the interactions}\label{sec:turningon}

Although it is possible to solve the general quench problem
between two interacting versions of the Luttinger model, we shall
focus here  on the cases where the interactions described by
$H_2$ and $H_4$ (cf. Eqs.~\ref{eq:hlm3} and \ref{eq:hlm4})
are suddenly switched on (this section),  and
switched  off (next section).  Thus, in this section, we shall assume that
 we have made the replacements
$g_{2,4}(q) \to g_{2,4}(q,t) = g_{2,4}(q) \theta(t)$ in Eqs.(\ref{eq:hlm3},\ref{eq:hlm4}).
The Hamiltonian at times $t > 0$
is therefore the interacting LM. In other words, using the notation introduced in
Sect.~\ref{sec:quadratic},
$H_{\rm f}  = H_{0} + H_2 + H_4 = H_{\rm LM}$, whereas the initial
Hamiltonian (for $t \leq 0$) is $H_{\rm i} = H_{0}$. However, we note that,
since both zero modes, $J$ and $N$, are conserved
 by $H_{\rm i} = H_{0}$ and $H_{\rm f} = H_{\rm LM}$,
their dynamics factors out, and
we shall  assume henceforth that we work within the sector of the Hilbert
space where $J = N = 0$ (this sector contains the non-interacting ground state, $|0\rangle$).
Therefore, from now on, we shall omit $H_\text{ZM}$ in all discussions.

As to the initial state, we shall consider that, within the
spirit of the sudden approximation, at $t=0$ the system is
prepared in a Boltzmann ensemble  at a temperature $T$
described by
\begin{equation}
\rho_0\equiv\rho(t=0)=Z_0^{-1}e^{-H_\text{i}/T}, \label{eq:init}
\end{equation}
where $Z_0=\Tr e^{-H_\text{i}/T}$ . We shall further assume that the
contact with the reservoir is removed at $t =0$, and that, after
the quench, the system undergoes a \emph{unitary} as it is
in isolation from the rest of the universe.

 Eq.~(\ref{eq:solution_time}) in the Appendix describes the solution to the
interaction quench in terms of the modes that annihilate the
initial ground state $|0\rangle$, the solution itself is not
particularly illuminating. To gain some insight into the
properties of the system following the quench, let us compute a
few observables. Amongst them, we  first turn our attention to
the instantaneous momentum distribution, which is the the
Fourier transform of the one-particle density matrix:
\begin{equation}\label{eq_Green_function}
C_{\psi_r}(x,t) = \langle e^{i H_\text{f} t/\hbar}
\psi^{\dag}_r(x) \psi^{\pd}_r(0) e^{-i H_\text{f} t/\hbar}
\rangle_0,
\end{equation}
where $\left\langle\cdots\right\rangle_0$ means that the
expectation value is taken over the ensemble described by $\rho_0$
(cf. Eq.~(\ref{eq:init})). The time dependence of the operators is
dictated by  $H_\text{f}$, as described in Section
\ref{sec:quadratic}. Notice that, since in general $[H_\text{f}, \rho_0]\neq
0$, time translation invariance is broken, and the
above correlation function is explicitly time-dependent.

The time evolution of  $\psi_\alpha(x)$ can be obtained
using the bosonization formula for the field
operator:\cite{luther_peschel_correlation_functions,%
haldane_luttinger_liquid,giamarchi_book_1d}
\begin{equation}\label{eq:bosonization_formula}
\psi_\alpha(x)=\frac{\eta_\alpha}{\sqrt{2\pi a}}\, e^{i
s_\alpha\phi_\alpha(x)},
\end{equation}
being  $\eta_r,\eta_l$  two  \emph{Majorana} operators (also known as
Klein factors) obeying $\{\eta_{\alpha},\eta_{\beta}\} = 2 \delta_{\alpha\beta}$, which
ensures the anticommutation of the left- and right-moving Fermi fields
(recall that $s_\alpha=+1$ for $\alpha=r$ and
$s_\alpha=-1$ for $\alpha=l$). The bosonic fields:
\begin{equation}\label{eq:boson_decomposition}
\phi_\alpha(x)=s_\alpha\varphi_{0\alpha}+\frac{2\pi x}{L}
N_\alpha+\Phi^\dag_\alpha(x)+\Phi^{\pd}_\alpha(x),
\end{equation}
where $[N_\alpha,\varphi_{0\beta}]=i\delta_{\alpha,\beta}$, and,
in terms of Fourier modes,
\begin{equation}\label{eq:boson_modes}
\Phi_\alpha(x) = \lim_{a_0\to 0^{+}} \sum_{q > 0} \left(
\frac{2\pi}{qL} \right)^{1/2} e^{-q a_0 /2} \, e^{i s_\alpha qx}
b(s_\alpha q).
\end{equation}
The details of the calculations of $C_{\psi_{r}}(x,t)$ have been
relegated to the Appendix~\ref{app:single_particle_correlations}.
In this section we will mainly describe the results. However, a number of
remarks about how the  calculations were performed
are in order before proceeding  any further. We first note that
interactions in the LM are assumed to be
long ranged.~\cite{luttinger_model,mattis_lieb_luttinger_model,haldane_luttinger_liquid}
This can be made explicit in the interaction couplings by writing $g_{2,4}(q) = g_{2,4}(qR_0)$,
where the length scale $R_0 \ll L$ is the interaction range. Thus, just
like system size $L$ plays the role of a cut-off for `infrared' (that is, long wave-length)
divergences, the interaction range, $R_0$
plays the role of an `ultra-violet'  cut-off that regulates
the short-distance divergences of the model. The results given below
were derived assuming a particular  form of the interaction (or
regularization scheme) where the  Bogoliubov parameter (cf. Eq.~\ref{eq:bogol_beta})
is chosen  such that $\sinh 2 \beta(q) = \gamma\, e^{-|q R_0|/2} $.
Furthermore, we replaced $v(q)$ by $v = v(0)$.
Indeed, these approximations are fairly
similar to  the ones used to compute the
time-dependent correlation
functions in equilibrium,\cite{luther_peschel_correlation_functions} given that
the expressions that we obtain for the out-equilibrium correlators are fairly
similar   to those of the equilibrium correlations.~\cite{luther_peschel_correlation_functions}
This regularization scheme greatly simplifies the calculations while not altering in a significant
way the asymptotic  behavior  of the correlators
for distances  much larger than $R_0$.\footnote{An exception  are pathological cases, like the Coulomb
interaction, where both $v(q)/|q|$ and $\sinh 2 \beta(q)$ are
singular for $q=0$.}

Returning to the one-particle density matrix
(cf. Eq.~\ref{eq_Green_function}), we note that it can be
written as  the product of two factors:
\begin{equation}
C_{\psi_r}(x,t)=C^{(0)}_{\psi_r}(x)h_r(x,t),
\end{equation}
where $C^{(0)}_{\psi_r}(x)$ is the
noninteracting one-particle density matrix, and thus $h_r(x,t)$ accounts for
deviations  due to the interactions. Hence, this factorization allows us
to obtain the instantaneous momentum distribution function as the
convolution:
\begin{equation}\label{eq:Fourier_transform_convolution}
f(p,t)=\int_{-\infty}^{\infty}\frac{dk}{2\pi}\,f^{(0)}(p-k)h_r(k,t),
\end{equation}
where $f^{(0)}(p)= (e^{\hbar v_F p/T} + 1)^{-1}$ is the
Fermi-Dirac distribution, and
\begin{equation}
h_r(k,t)  = \int^{+\infty}_{-\infty} dx \, e^{ipx} h_r(x,t).
\end{equation}

Before presenting the results for the expression of  the one-particle
density matrix as well as the
momentum distribution at finite temperatures,  it is worth
considering the much simpler looking  zero-temperature
expression. We first discuss finite-size effects.
For a system of size $L$ we obtain:\cite{cazalilla_quench_LL}
\begin{multline}
C_{\psi_r}(x|L)=C^{(0)}_{\psi_r}(x|L)\left\vert\frac{R_0}{d(x|L)}\right\vert^{\gamma^2}\\
\times\left\vert\frac{d(x-2vt|L)d(x+2vt|L)}{d(2vt|L)d(-2vt|L)}\right\vert^{\gamma^2/2},\label{eq:green_function_LM}
\end{multline}
where $d(z|L)=L|\sin(\pi z/L)|$ is the \emph{cord} function and
$C^{(0)}_{\psi_r}(x|L)=i \left\{2L\sin
\left[\pi(x+ia_0)/L\right]\right\}^{-1}$ ($a_0\to 0^{+}$) the
noninteracting one-particle density matrix. Notice that this
result is valid only asymptotically, that is, for $d(x|L), d(x\pm
2 vt) \gg R_0$. Thus, we see that $C_{\psi_r}(x,t|L)$ is a
periodic function of time with period equal to $\tau_0 = L/2v$. This is in
agreement with the general expectation that correlations in finite-size
systems exhibit time recurrences because the energy spectrum is discrete. Although
the recurrence time generally depends on the details of the energy spectrum,
in the LM  the spectrum is linear $\omega(q) \simeq v |q|$ and therefore
the energy spacing between (non-degenerate) many-body states is $\Delta_0 \simeq 2\pi \hbar v/L$.
Hence, the recurrence time $\tau_0 \sim 2\pi / \Delta_0$ follows (the extra factor of $\frac{1}{2}$ is explained
by the so-called light-cone effect, see further below).  The recurrent behavior exhibited by
the one-particle density matrix~(\ref{eq_Green_function}) implies that, after the quench,
the system does not reach a time-independent stationary state as time grows.
 A similar conclusion is reached by analyzing  other correlations, such as
\emph{e.g.} the  finite-size version of the
density correlation function,\cite{cazalilla_quench_LL}
\begin{align}
&C_{J_r}(x,t|L) = \langle e^{i H_\text{f} t/\hbar} J_r(x) J_r(0)
e^{-i H_\text{f} t/\hbar} \rangle_0
\nonumber\\
&= - \frac{(1+\gamma^2)/4\pi^2}{\left[d(x|L)\right]^2}
+ \frac{\gamma^2/8\pi^2}{\left[ d(x-2vt|L)\right]^2}
 + \frac{\gamma^2/8\pi^2}{\left[ d(x+2vt|L)\right]^2},\nonumber\\
 \label{eq:dc}
\end{align}
where
\begin{equation}
J_{r}(x) =\sum_{q}  \frac{e^{i q x}}{L}  \: J_{r}(q) = :\psi^{\dag}_{r}(x)\psi_r(x):\, = \frac{1}{2\pi} \partial_x \phi_r(x),
\label{eq:currents}
\end{equation}
is the density (also referred to as 'current') operator in real space.

In the thermodynamic limit, $L\to\infty$, the recurrence time
$\tau_0 = L/2v \to +\infty$, and the system does indeed reach a time-independent
steady state.  In this limit,  $d(x|L)\to |x|$, and the single particle density matrix
becomes:\cite{cazalilla_quench_LL}
\begin{equation}
C_{\psi_r}(x,t>0)=G_r^{(0)}(x)\left|\frac{R_0}{x}\right|^{\gamma^2}
\left|\frac{x^2 - (2vt)^2}{(2vt)^2}\right|^{\gamma^2/2}. \label{eq:cpsi}
\end{equation}
Hence,
\begin{equation}
h_r(x,t) = \left|\frac{R_0}{x}\right|^{\gamma^2}
\left|\frac{x^2 - (2vt)^2}{(2vt)^2}\right|^{\gamma^2/2}.
\end{equation}
In order to understand the evolution of the momentum distribution, without actually
having to compute it,  it  useful to consider the various limits of the above
expression, Eq.~(\ref{eq:cpsi}). First of all, for short times  such that
$2vt\ll |x|$, the function $h_r(x,t)$ is
asymptotically just a time-dependent factor,\cite{cazalilla_quench_LL}
\begin{equation}
h_r(x,t)\simeq
Z(t)=\left(\frac{R_0}{2vt}\right)^{\gamma^2},
\end{equation}
which can be interpreted as a time-dependent  `Landau quasi-particle'
weight in an effective (time-dependent)  Fermi-liquid
description of the system. In other words, as time evolves after
the quench, we could imagine that the quasi-particle weight at the Fermi level is
reduced from its initial value, $Z(t=0) = 1$, to $Z(t > 0) < 1$.
At zero temperature, this time-dependent renormalization of the
quasi-particle weight reflects itself in a reduction of the
discontinuity of the momentum
distribution $f(p,t)$ at the Fermi level (which is located at
$p = 0$ in our convention). Therefore, at any finite time, the system behaves
as if it was a Fermi liquid and therefore it keeps memory of the initial state, that is, a
non-interacting Fermi gas.

\begin{center}
\begin{figure}
\includegraphics[width=\largefigwidth]{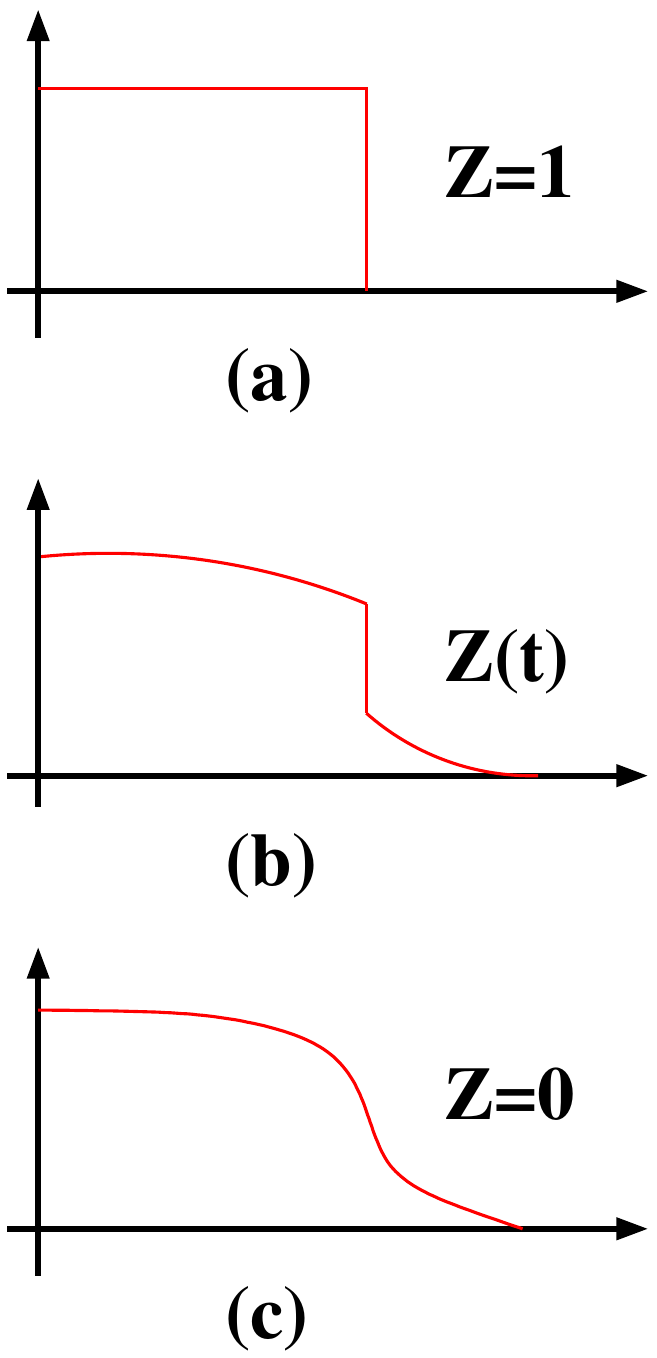}
\caption{Schematic time evolution of the momentum distribution $f(p,t)$ at
zero temperature.  (a) At $t=0$, the momentum distribution is that
of non-interacting fermions, with a  discontinuity at the Fermi
level ($p = 0$) $Z=1$. (b) At $t >0$ the discontinuity is reduced
in a power-law fashion $Z(t) \sim t^{-\gamma^2}$. (c) For $t \to
+\infty$ the discontinuity disappears and the momentum
distribution exhibits a power-law singularity close to the Fermi
level $p = 0$, $f(p,t\to +\infty) = \frac{1}{2} - \text{const.}\times
|p|^{\gamma^2}$. However, the exponent characterizing the singularity is
not the equilibrium exponent.}\label{fig:fpt}
\end{figure}
\end{center}

Yet, for $t\to +\infty$,
$h_r(x,t)$ becomes a power-law:~\cite{cazalilla_quench_LL}
\begin{equation}\label{eq:long_times_correlator_zero_T}
\lim_{t\to\infty}
h_r(x ,t)=\left|\frac{R_0}{x}\right|^{\gamma^2}.
\end{equation}
and also does the momentum distribution. The behavior of
the momentum distribution at different times is depicted in
Fig.~\ref{fig:fpt}. Interestingly, this time-dependent reduction of the quasi-particle weight after being quenched
into the interacting state has been also found in
Ref.~\onlinecite{kehrein_quench_hubbard} when studying an interaction quench
in the Hubbard model in the limit of infinite dimensions. In this case, however,
the discontinuity remains finite even for $t \to +\infty$, which is different
from the behavior of the LM, which is known to be a non-Fermi
liquid at equilibrium.  Some of these non-fermi-liquid features also persist in the quench
dynamics.

The result of Eq.~(\ref{eq:long_times_correlator_zero_T})
is similar to the zero temperature result in equilibrium. However, the
exponent of $C_{\psi_r}(x,t\to \infty)$ is equal to $1+ \gamma^2$,
and, even for an infinitesimal interaction  (\emph{i.e.} $\gamma \ll
1$), it is always larger than the one that governs the  ground state (\emph{i.e.} equilibrium)
correlations:\cite{luther_peschel_correlation_functions,haldane_luttinger_liquid}
$\gamma^2  = \sinh^2 2\beta(0) >
\gamma_\text{eq}^2=2\sinh^2\beta(0)$. The reason for the larger exponent
can be qualitatively understood from the following facts: \emph{i}) Because of
the variational theorem, the initial state
(\emph{i.e.} the ground state of the non-interacting Hamiltonian $H_\mathrm{i} = H_0$), is a
complicated excited state of the final Hamiltonian $H_\mathrm{f} = H_{\rm LM}$. \emph{ii}) Both $H_\mathrm{i}$  and
$H_{\rm f}$ are critical (\emph{i.e.} scale free, apart from the cut-off $R_0$),
thus, the system is likely to remain critical. To end the present discussion,
we note that, in the literature on Tomonaga-Luttinger liquids,
it  is customary to introduce the dimensionless parameter $K = e^{-2\beta(0)}$,
in terms of which $\gamma^2 = (K-K^{-1})^2/4$, which needs to be compared with the equilibrium
exponent resulting from $\gamma_\text{eq} =  (K + K^{-1} - 2)/2$.
We shall  use the parameter $K$ in other expressions below.

It is worth emphasizing that the particular evolution of the asymptotic
correlations from Fermi liquid-like at short times to  non-Fermi liquid-like at infinite time
exhibited by the one-particle density matrix, is also found in other correlation functions.
However, the idea that the system 'looks like' an interacting Fermi liquid at any finite $t$
should not be taken too far. In this regard, we should note that
for  $|x| \gg 2 vt$,  the prefactor of the term $\propto (2\pi x)^{-2}$
of the density correlation function (cf. Eq~(\ref{eq:dc})),
which in equilibrium is proportional to the system compressibility,~\cite{giamarchi_book_1d}
remains equal to (minus) unity, which is the value that corresponds
to a non-interacting Fermi gas (in an
interacting Fermi liquid it would deviate from one).
For $t \to +\infty$ the same prefactor becomes $(1 + \gamma^2) > 1$, which does not
relate easily to a non-Fermi liquid-like behavior. However, other correlation
functions (like the one-particle density matrix discussed above) exhibit a similar
behavior to the one-particle density matrix.  For instance, let us consider
the following correlators:
\begin{align}
C^m_{\phi}(x,t) &=  \langle e^{2i m \phi(x,t)} e^{-2im\phi(0,t)} \rangle, \label{eq:corr_phi} \\
C^n_{\theta}(x,t) &= \langle e^{in\theta(x,t)} e^{-i n \theta(0,t)} \rangle, \label{eq:corr_theta}
\end{align}
where $\phi(x) = \frac{1}{2}\left[\phi_r(x) + \phi_l(x) \right]$ and $\theta(x) =
\frac{1}{2}\left[\phi_r(x) - \phi_l(x)\right]$,
the spatial derivatives  of $\phi$ and $\theta$ are related to the (total)
density and current density fluctuations,
respectively.\footnote{In a Luttinger liquid, the $C^m_{\phi}$ correlator
describes the fluctuations of wave number close to
$2m k_F$ (where $k_F$ is the Fermi momentum) of  the density-correlation
function~\cite{haldane_luttinger_liquid,giamarchi_book_1d}. In the presence of a periodic potential
of periodicity equal to $2 m k_F$ the system may become an insulator.~\cite{giamarchi_book_1d}
The power-law behavior exhibited  at zero temperature, and in the thermodynamic limit)
is a consequence  of the gapless spectrum and  the absence of long
range order in the density. In the insulating (\emph{i.e.} gapped) phase, this correlation
function decays to a non-zero constant at long distances, which is a consequence of the existence
of long-ranged order in the density at wave  number $2 m k_F$.  Similarly, in equilibrium $C^n_{\theta}$
measures the phase fluctuations, and exhibits
a power-law, reflecting the absence of long range order in the phase. However, in the Luttinger model of interest for us here,
terms with $m > 0$ are absent from the density operator, which is given by
$\rho(x) = J_r(x) + J_l(x)$.~\cite{haldane_luttinger_liquid}}
Using exactly the same methods as above, we
find (for $L \to \infty$):
\begin{align}
\frac{C^m_{\phi} (x,t)}{C^{(0,m)}_{\phi}(x)} =
\left|\left( \frac{R_0}{2 vt } \right)^2 \frac{x^2 - (2vt)^2}{x^2} \right|^{m^2 (K^2 - 1)/2}, \\
\frac{C^n_{\phi} (x,t)}{C^{(0,n)}_{\phi}(x)} =
\left|\left( \frac{R_0}{2 vt } \right)^2 \frac{x^2 - (2vt)^2}{x^2} \right|^{n^2 (K^{-2} - 1)/8},
\end{align}
where $C^{(0,m)}(x) =  A^{\phi}_m |R_0/x|^{2m^2}$ and $C^{(0,n)}(x) = A^{\theta}_n |R_0/x|^{n^2/2}$
are the non-interacting correlation function (where $A^{\phi}_m$ and $A^{\theta}_{n}$ are
non-universal  prefactors).  We note that the usual duality relation where $\phi \to \theta$ and
$K \to K^{-1}$, which one encounters when studying equilibrium correlation functions,~\cite{giamarchi_book_1d}
still holds for these non-equilibrium correlators.  Let us next analyze their
asymptotic properties. We consider only $C^{m}_{\phi}(x,t)$, as identical conclusions also
apply to $C^{n}_{\theta}(x,t)$ by virtue of the duality relation. For $|x| \gg 2 v t$,  we have:
\begin{align}
C^{m}_{\phi}(x,t)=  C^{(0,m)}(x)\ \left(\frac{R_0}{2 v t} \right)^{m^2 (K^2-1)}.
\end{align}
Thus, up to the time-dependent pre-factor,
correlations take the form of a non-interacting system of Fermions, $C^{(0,m)}(x)$.
However, in the opposite limit  ($|x| \ll 2 v t$),  this correlator exhibits a
non-trivial power-law:
\begin{align}
C^m_{\theta}(x,t) \simeq   \left|\frac{R_0}{x} \right|^{m^2(K^2+1)}. \label{eq:long_times_2kf}
\end{align}
Notice that this expression also describes   the infinite-time behavior, which
is controlled by an exponent equal to   $m^2(K^2+1)$, being again different
from the exponent exhibited by the
same correlator in equilibrium, which equals $2 m^2 \left[\cosh 2 \beta(0) -
\sinh 2\beta(0) \right] = 2 m^2 K$.

In order to understand why the behavior found for $t \to \infty$ in
Eqs.~(\ref{eq:long_times_correlator_zero_T},\ref{eq:long_times_2kf})
also holds for  $|x| \ll 2 vt$, let us
consider the initial state at zero temperature,
$\rho_0 =|0\rangle \langle 0|$.~\cite{calabrese_quench_CFT,calabrese_quench_CFT_long}
As mentioned above, this is a rather complicated excited state of the Hamiltonian that
performs the time-evolution, $H_\text{f} = H_{\rm LM}$. This means that,
initially, there are a large number of excitations of
$H_\text{f}$, namely, phonons with dispersion $\omega(q) = v(q)
|q|$. The distribution of the phonons $\langle b^{\dag}(q) b(q)
\rangle_0 = \sinh^2 \beta(q)$ is time-independent and peaked at $q = 0$. Thus,
within the approximation where $v(q) \simeq v(0) = v$,  the
excitations  propagate between two given  points with  velocity
$v$. Thus, if we consider the
correlations at two points $A$ and $B$ separated by a distance $|x|$,
the nature of the correlation at a give time $t$ depends on wether the excitations
found initially at, say,  point $A$,  have  been able to reach point $B$
or not. This is not the case if $|x| >  2
vt$, and thus correlations retain essentially the properties they had
in the initial state. Thus, up to a time-dependent prefactor,
$C_{\psi_r}(x,t) \propto C^{(0)}_{\psi_r}(x)$. However, if the
two points have been able to `talk to each other' through the
excitations present in the initial state, then correlations will
be qualitatively different. This happens for a time $t = t_0$ when the
excitations propagating from $A$ meet the excitations propagating from $B$,
that is, for $x - vt_0 = vt_0$, or $t_0 = x/2v$ (we assume $x > 0$ without
loss of generality). Thus, for
given separation $x$ and time $t$, there is a length scale $2 vt$, which
marks the transition between two different regimes in the
correlations. In the instantaneous momentum distribution, this
reflects itself in a crossover as a function of time
from a momentum distribution $n(p)$ exhibiting a discontinuous
Fermi liquid-like behavior,  which is valid
\emph{i.e.} for $|p| \ll (2vt)^{-1}$) to a power-law
behavior of the form $ \sim |p R_0|^{\gamma^2} {\rm sgn}(p)$, which
applies for $|p| \gg (2
vt)^{-1}$ but $|p| \ll R^{-1}_0$ (for $|p| \gg R^{-1}_0$ we
recover the free particle behavior corresponding to the Fermi-Dirac
distribution function at $T = 0$). In the $t \to \infty$ limit, by
using the regularization scheme described above, the asymptotic
momentum distribution at zero temperature  can be
obtained  with the help of tables.\cite{gradshteyn80_tables} The resulting formula behaves as the
non-interacting Fermi-Dirac distribution for $|p| \gg R^{-1}_0$, whereas
for $|p| \gg R^{-1}_{0}$, it describes a non-Fermi liquid-like steady state:
\begin{multline}
f(p, t \to +\infty)=\frac{1}{2}-\frac{pR_0}{2}
\Big[K_{\frac{\gamma^2-1}{2}}(|p R_0|){\cal L}_{\frac{\gamma^2-3}{2}}(|pR_0|)\\
+K_{\frac{\gamma^2-3}{2}}(|pR_0|){\cal
L}_{\frac{\gamma^2-1}{2}}(|p R_0|)\Big],
\end{multline}
where $K_{\nu}(z)$,   ${\cal L}_{\nu}(z)$ are the modified Bessel
and Struve functions,\cite{gradshteyn80_tables} respectively. This
expression yields a power law for $|p R_0|\ll 1$, where
$n(p,t\to+\infty)\simeq \frac{1}{2}-\text{const.}\times
(|pR_0|)^{\gamma^2} \, \sign(p)$. Note that the momentum
distribution $n(p =0,t) = \frac{1}{2}$, which is given by the
invariance of the LM under  particle-hole symmetry
$\psi_{\alpha}(p) \to \psi^{\dag}_{\alpha}(-p)$.

\begin{center}
\begin{figure}
\includegraphics[width=\figwidth]{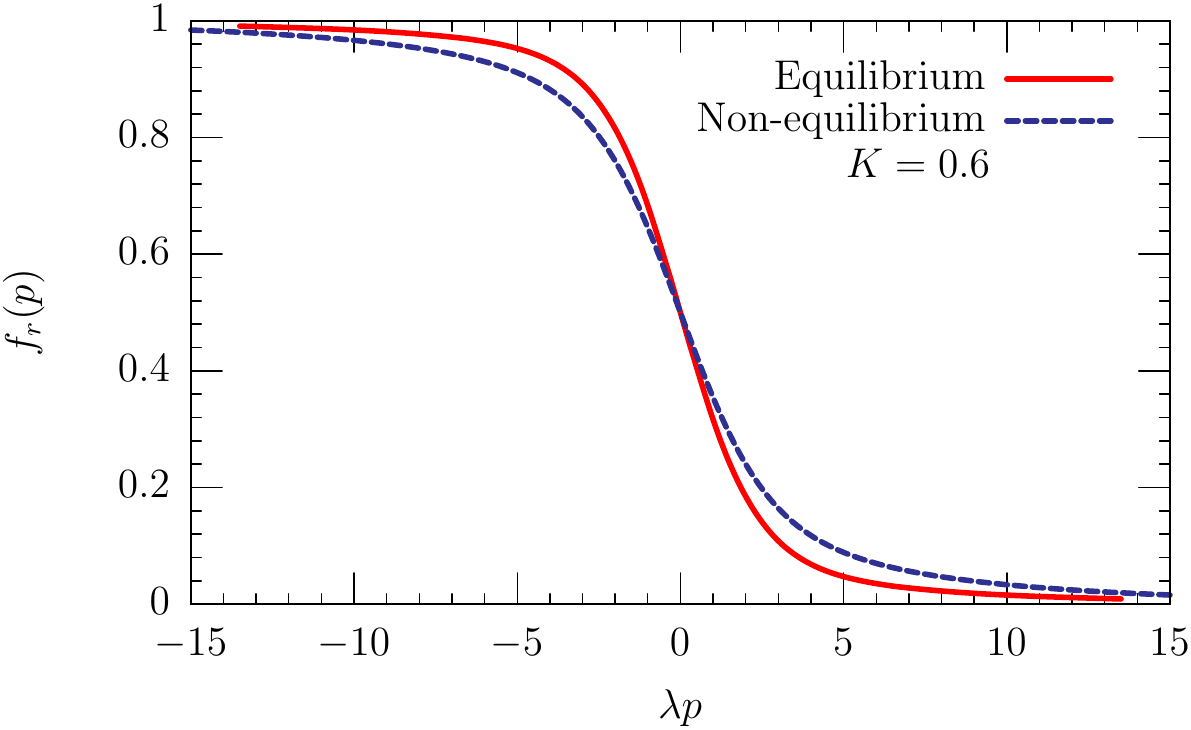}
\caption{Infinite time limit of the  momentum distribution vs.
$\lambda p$ ($\lambda = \hbar v_F /T$ is the
thermal correlation length in the initial state) for a non-interacting Luttinger model at finite
temperature $T$ that is quenched  into an interacting state with repulsive interactions
(corresponding to a Luttinger parameter $K=0.6$).}\label{fig:mom_dist_small}
\end{figure}
\end{center}

\begin{center}
\begin{figure}
\includegraphics[width=\figwidth]{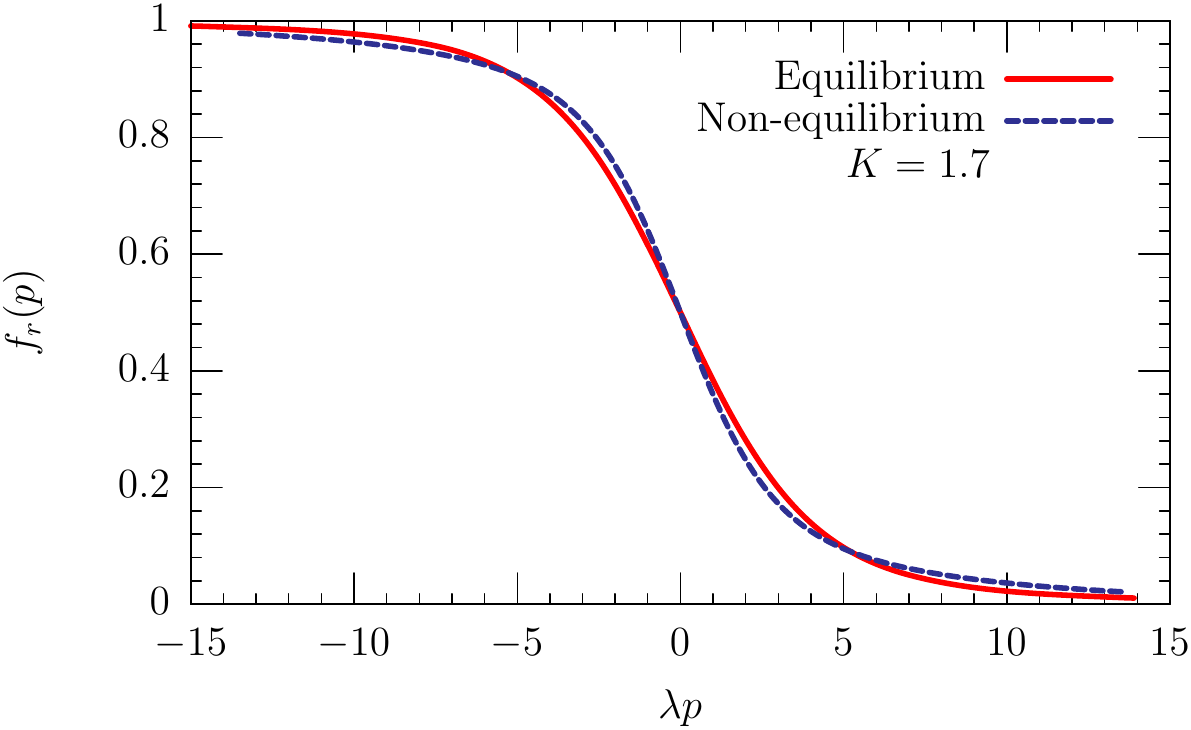}
\caption{Infinite time limit of the  momentum distribution vs.
$\lambda p$ ($\lambda = \hbar v_F /T$ is the
thermal correlation length in the initial state) for a non-interacting Luttinger model at finite
temperature $T$ that is quenched  into an interacting state with attractive interactions
(corresponding to a Luttinger parameter $K=1.7$).}\label{fig:mom_dist_large}
\end{figure}
\end{center}

Let us finally present  the generalization
of the above results for the one-particle density matrix
to finite temperatures, $T > 0$. For $T \ll \hbar v_F R_0$ (but $T \gg \Delta_0 = 2\pi \hbar v/L$,
so that we can neglect finite-size effects and effectively take
the thermodynamic limit) $C_{\psi_r}(x,t)$ takes the
following asymptotic  form:
\begin{multline}
C_{\psi_r}(x,t>0|T)=C_{\psi_r}^{(0)}(x|T)\left\vert\frac{\pi  R_0/\lambda}
{dh(x|T)}\right\vert^{\gamma^2}\\
\times\left\vert\frac{dh(x-2vt|T)dh(x+2vt|T)}{dh(2vt|T)dh(-2vt|T)}
\right\vert^{\gamma^2/2},\label{eq:corr_func_fin_T}
\end{multline}
where $C^{(0)}_{\psi_r}(x|T)$ and $dh(x|T)$ can be obtained from
$C^{(0)}_{\psi_r}(x|L)$ and $d(x|L)$ by replacing $L\sin(\pi x/L)/\pi$ by
$\lambda \sinh(\pi x/\lambda)$, where $\lambda=\hbar v_F/T$ is the
thermal correlation length. At long times, $h_r(x,t|T)$ reduces to
\begin{equation}
h_r(x)=\left| \frac{\pi R_0/\lambda}{ \sinh
\left( \pi x/ \lambda \right)}
\right|^{\gamma^2}.\label{eq:factor_h_fin_T}
\end{equation}
Therefore we again find that $C_{\psi_r}(x,t\to\infty|T)$ has a form similar to the
the equilibrium correlation function at finite temperature with a different exponent
controlling the asymptotic exponential decay of correlations. Notice that the exponential
decay the correlations for $t > 0$ is a direct consequence of the fact that the
initial state has a characteristic correlation length, the thermal
correlation length $\lambda=\hbar v_F/T$.\footnote{A similar situation is found when analyzing quenches at $T = 0$ from a non-critical (that is, gapped) into a critical (that is, gapless) state. In that case, the role of $\lambda$ will be played by the correlation length of the system that is determined by the (inverse of the) energy gap in the initial state.\cite{calabrese_quench_CFT,unpub}}
The exponential decay of correlations at finite $T$ implies  that the the steady state will be reached
exponentially rapidly in a time of the order of $\hbar/T$.
It is also worth noting, however, that  the above expression depends parametrically on the
thermal correlation length $\lambda$, and the Fermi velocity,  $v_F$, which enters in the
expression for $\lambda =  \hbar v_F/T$,
instead of the (renormalized) phonon velocity which enters in the thermal length
$\lambda_{\rm eq} = \hbar v/T$, characterizing equilibrium correlations.
Thus, since the velocity appears
only through the definition of the thermal correlation length
$\lambda$, or, in other words, in combination with the
temperature, the change from $v$ to $v_F$ can be also understood as
an change in the  temperature scale. Furthermore, in a system with Galilean
symmetry,~\cite{giamarchi_book_1d} we have that  $v K=v_F$ and thus the parameter that
controls the temperature scale now is the Luttinger parameter $K$,
so that the  asymptotic correlations at $t\to \infty$ can be regarded as the equilibrium
correlations with a different exponent and an effective temperature, $T_\text{eff}=T/K$.
Thus, for repulsive interactions (\emph{i.e.} $K<1$) we could say that, besides modifying the exponent,
the interaction quench increases the effective temperature,
whereas for attractive interactions (\emph{i.e.} $K>1$) the effective
temperature is reduced after the quench. This effect has an impact on
the momentum distribution at finite temperatures because it compensates
the effect of the larger non-equilibrium exponent
on the momentum distribution at finite $T$. To show this,
we need to obtain the Fourier transform of
$h_r(x)$. This can be  done by relating it to an integral
representation of the associated Legendre function $P_\mu^\nu(z)$,
\cite{gradshteyn80_tables} and thus the Fourier transform of $h_r(x)$
can be written as:
\begin{multline}
h_r(p)=\frac{\lambda}{\sqrt{\pi}}\left(\frac{\pi
R_0}{\lambda}\right)^{(\gamma^2+1)/2}
\frac{\left|\Gamma\left(\frac{\gamma^2}{2}+\frac{i\lambda
p}{2\pi}\right)\right|^2}{\Gamma\left(\frac{\gamma^2}{2}\right)} \\
\times P_{\frac{i\lambda
p}{2\pi}-\frac{1}{2}}^{-\frac{\gamma^2}{2}+\frac{1}{2}}\left[ -\cos \left( \frac{2\pi
R_0}{\lambda}\right)\right].\label{eq:h_factor_finite_T}
\end{multline}
Hence, the momentum distribution can be obtained by numerically evaluating
the convolution with the Fermi-Dirac distribution function (cf. Eq.~\ref{eq:Fourier_transform_convolution})
of the above expression,  Eq.~(\ref{eq:h_factor_finite_T}).
In Figs. \ref{fig:mom_dist_small} and \ref{fig:mom_dist_large}
the momentum distribution of the interacting system for $t \gg \hbar/T$
is displayed for  a non-interacting LM that undergoes an interaction quench
with repulsive (corresponding to $K=0.6$) and attractive (corresponding to $K=1.7$) interactions,
respectively.
\subsection{Suddenly turning-off the interactions}

Next we  briefly consider the opposite situation to the one analyzed above,
namely  the case where interaction between the fermions in the initial state  suddenly
disappears.   The fact that the initial state is a
highly complicated state of the Hamiltonian that performs the time evolution (in this case
$H_{\rm f} = H_0$,  cf. Eq.~\ref{eq:hlm1}), implies that we cannot expect that a Fermi liquid
will emerge asymptotically at long times after the quench. Indeed, at zero temperature
a thermodynamically large system
approaches a steady state exhibiting equal-time correlations that decay algebraically in space.
However, the exponents differ again from the (non-interacting) equilibrium ones. This can
be illustrated by, \emph{e.g.} computing the following correlation functions:
\begin{align}
C^m_{\phi}(x,t)  &=  \langle e^{2im \theta(x,t)} e^{-2i m \theta(0,t)} \rangle \nonumber\\
&=  I^{m}_{\phi}(x) \: \left( \frac{R_0}{2 v_F t}\right)^{m^2(K^{-1}-K)} \nonumber\\
& \quad \times \left|\frac{x^2 - (2v_Ft)^2}{x^2}\right|^{m^2(K^{-1}-K)/2},  \\
C^n_{\theta}(x,t) &= \langle e^{in \phi(x,t)} e^{-i n \phi(0,t)} \rangle \nonumber\\
&=  I^n_{\theta}(x)  \left( \frac{R_0}{2 v_F t}\right)^{n^2(K-K^{-1})/2}
\nonumber\\ & \quad \times
\left|\frac{x^2 - (2v_Ft)^2}{x^2}\right|^{n^2(K-K^{-1})/4},
\end{align}
where
\begin{align}
I^m_{\phi}(x) &= \left|\frac{R_0}{x}\right|^{2m^2 K},\\
I^n_{\theta}(x) &=  \left|\frac{R_0}{x}\right|^{n^2/2K}
\end{align}
are the correlation functions in the initial (interacting) ground state ($A_{\theta/\phi}$
are non-universal prefactors).   We note again that the duality $\theta \to \phi$ and $K \to K^{-1}$ also
holds in this case. The correlations in the stationary state that is asymptotically approached
at long times read:
\begin{align}
\lim_{t \to +\infty} C^m_{\theta}(x,t)  &=  A_{\theta} \left| \frac{R_0}{x} \right|^{m^2(K^{-1}+K)}, \\
\lim_{t \to +\infty} C^n_{\phi}(x,t)  &=  A_{\phi} \left| \frac{R_0}{x} \right|^{n^2(K^{-1}+K)/2}.
\end{align}
However, at short times, $t \ll |x|/2v_F$, correlations look like those of in the initial state,
up to a time-dependent prefactor:
\begin{align}
C^m_{\theta}(x,t \ll |x|/2v_F )  &=  \left( \frac{R_0}{2 v_F t}\right)^{m^2(K^{-1}-K)} \:  I^{m}_{\theta}(x)   \\
C^n_{\phi}(x,t \ll |x|/2v_F)  &=     \left( \frac{R_0}{2 v_F t}\right)^{n^2(K-K^{-1})/2} \:  I^n_{\phi}(x).
\end{align}
In this case the time-dependent prefactor has also a power-law form.

\section{Long-time dynamics and the generalized Gibbs ensemble}
\label{sec:generalized}

Recently, Rigol and coworkers~\cite{rigol_generalized_gibbs_hcbosons}
observed that, at least for observables like the momentum distribution or
the ground state density, their long-time behavior following a quantum quench
in an integrable system can described by adopting the maximum entropy (also
called `subjective') approach to Statistical Mechanics
pioneered by Jaynes.~\cite{JaynesI,JaynesII,balian_stat_mech}
Within this approach, the equilibrium state of a system is
described by  a density matrix
that extremizes the von-Neumann entropy, $S=-\Tr\rho\ln\rho$,
subject to the constraints provided  by a certain set of  integrals of motion of the system.
In the case of an integrable system, if  $\{I_{m}\}$ is a set of certain (but not all of the possible)
independent integrals of motion of the system,  this procedure leads to a `generalized' Gibbs ensemble,
described by the following density matrix:
\begin{equation}
\rho_{\rm gG} = \frac{1}{Z_{\rm gG}} e^{-\sum_m \lambda_m
I_m},\label{eq:density_operator_gG}
\end{equation}
where  $Z_{\rm gG} = \Tr e^{-\sum_m \lambda_m I_m}$. The values of the Lagrange
multipliers  $\lambda_m$ must be determined from the condition that
\begin{equation}
\langle I_m \rangle_\text{gG} =  {\rm Tr} \left[ \rho_{0} I_m  \right]
=  \langle I_m \rangle.
\label{eq:inic_cond_T0}
\end{equation}
where $\rho_{0}$
describes the initial state of the system, and $\langle \cdots \rangle_\text{gG}$ stands for  the average
taken over the generalized Gibbs ensemble,  Eq.
(\ref{eq:density_operator_gG}). Although  $\rho_i = |\Phi(t = 0) \rangle \langle \Phi(t= 0)|$
in the case of a pure state, as was first used in Ref.~\onlinecite{rigol_generalized_gibbs_hcbosons},
nothing prevent us from taking $\rho_{0}$ to be an arbitrary mixed state  and in particular
a thermal state characterized by an absolute temperature $T$.  In such a case,  the
 Lagrange multipliers will  depend on $T$ or any other parameter that defines the initial state.

Rigol and coworkers tested  numerically the above
conjecture by studying the quench dynamics of a 1D lattice gas of hard-core bosons
(see Ref.~\onlinecite{rigol_generalized_gibbs_hcbosons,Rigol_hc_noneq2} for
more details). The question that naturally arises then  is whether correlations of the LM also relax
in agreement with the above conjecture. In other words, does the average
$\langle O\rangle(t)$  at long times relax to the value
$\langle O\rangle_\text{gG} = {\rm Tr}\:  \rho_{\rm gG} \, O$, for any
of the correlation functions considered previously?
We shall first discuss an observable for which the generalized Gibbs
ensemble fails to reproduce their expectation values. Moreover,
by considering the correlation function of the
current operators (\emph{i.e.} $O = J_r(x)$, cf. Eq~\ref{eq:currents}),
we will illustrate why it works in the LM. Calculations of other observables can be found in the appendices.

Let us define the generalized Gibbs ensemble for the
Luttinger model (LM). Since the final Hamiltonian (in the $N = 0$ and $J = 0$ sector)
is diagonal in the $b$ boson operator basis, \emph{i.e.} $H_{LM} = \sum_{q\neq 0}
\hbar v(q) |q| \: b^{\dag}(q) b(q)$, a natural choice for the set of
integrals of motion is $I_m \to I(q)  = n(q) = b^{\dag}(q) b(q)$
for all $q \neq 0$ (a more complete version of the ensemble should also include $N$ and $J$,
but this will not be necessary as we work in the thermodynamic limit).
Thus, for the quench from the non-interacting to the interacting state
(cf. Sect.~\ref{sec:turningon}), where the initial state is
$|\Phi(t =0) \rangle = |0\rangle$,  the Lagrange multipliers
are  determined by Eq.~(\ref{eq:inic_cond_T0}), which yields:
\begin{equation}\label{eq:inic_condition}
 \langle I(q) \rangle_{\text{gG}} =
 \langle n(q) \rangle_{\text{gG}} = \sinh^2\beta(q)=\frac{1}{e^{\lambda(q)}-1}.
\end{equation}
Indeed, this result can be quickly established by realizing that $\rho_{\text{gG}}$
has the same form as the density matrix of a peculiar
canonical ensemble where the temperature on each eigenmode of the final Hamiltonian
depends on the wave-vector  $q$,
that is, $T(q)=\hbar v(q)|q|/\lambda(q)$. Alternatively,
one can also regard it as an ensemble where the effective Hamiltonian that defines the
Boltzmann weight is given by $H_\text{eff}/T_\text{eff}=\sum_{q\neq 0} \lambda(q)
n(q)$. However, it is worth noting that $\rho_{\rm gG}$
is diagonal in $n(q)$, and therefore it does not capture the correlations existing
in the initial state between the $q$ and $-q$ modes. Mathematically,
\begin{multline}
\langle n(q) n(-q) \rangle = \sinh^2 \beta(q) \cosh 2 \beta(q)
\neq \langle n(q) n(-q) \rangle_{\text{gG}}  \\
=  \langle n(q) \rangle_{\text{gG}}  \langle n(-q) \rangle_{\text{gG}}
= \sinh^4 \beta(q)\,.
\end{multline}
As matter of fact, since $n(q) n(-q)$ commutes with $H$, we
conclude from the above that $\langle n(q) n(-q) \rangle$ does not relax to the
value predicted by $\rho_{\rm gG}$. Although this defect of
$\rho_{\rm gG}$ can be fixed by enlarging the set of integrals
of motion to include the set $I'(q) = n(q) n(-q)$ as well, and the corresponding lagrange multipliers,
we shall show below that this is not needed. The reason is that the correlations missing in the
generalized Gibbs ensemble as defined above yield a vanishing contribution in the thermodynamic
limit to the simplest correlation functions. However, before discussing this
point, it is worth mentioning one important exception to the class of observables that
relax according to the generalized Gibbs ensemble, namely
the squared fluctuations of the energy:
\begin{multline}
\sigma^2=\langle H^2 \rangle - \langle H \rangle^2 =
\sum_{p,q}\hbar\omega(p)\hbar\omega(q)\\
\times[\langle n(p)n(q)\rangle-\langle n(p)\rangle\langle n(q)\rangle]
\end{multline}
which yields  $\sigma^2 =2\sigma^2_\text{gG}=\sum_q\sinh^2
2\beta(q)\hbar^2\omega(q)^2$.  Again, since the operator
$H^2$ is conserved, $\sigma^2$ violates the relaxation
hypothesis. However, it is  tempting to argue since $\sigma^2$
(like $\langle H \rangle$) is a non-universal
property of the LM model, this violation is less problematic
than a violation in the asymptotic behavior of the correlation functions would be,
as the latter tends to be more universal.

 In order to proceed with the discussion of the validity of the
generalized Gibbs ensemble, let us consider, for the sake of simplicity, the correlations the current
operator. In particular, we shall study the following two-time correlation function
(no time ordering is implied):
\begin{equation}
C_{J_r}(x,t,\tau)=\left\langle J_r(x,t+\tau/2)J_r(0,t-\tau/2)\right\rangle_T.
\end{equation}
where $\langle \ldots \rangle_T$ stands for average over the
initial thermal ensemble described by $\rho_i = e^{-H_{\text{i}}/T}/Z_0$,
with $H_{\rm i} = H_0$ (cf. Eq.~\ref{eq:hlm1}). Using
(\ref{eq:boson_decomposition}),   and (\ref{eq:boson_modes}), we obtain:
\begin{align}
&C_{J_r}(x,t,\tau)=\frac{1}{(2\pi)^2}\sum_{q>0}\left(
\frac{2\pi q}{L}\right)\: e^{-qa_0}\notag\\
\times&\Big\{e^{iqx} f(q,t+\tau/2)f^\ast(q,t-\tau/2)\left[ 1+n_B(q,T)\right]\notag\\
+&e^{iqx}g^\ast(q,t + \tau/2)g(q,t - \tau/2)n_B(q,T)\notag\\
+&e^{-iqx}f^\ast(q,t + \tau/2)f(q,t-\tau/2)n_B(q,T)\notag\\
+&e^{-iqx}g(q,t+\tau/2)g^\ast(q,t-\tau/2)\left[  1+n_B(q,T)
\right]\Big\},\label{eq:two_points_can}
\end{align}
being $n_B(q,T) = \langle a^{\dag}(q) a(q) \rangle = (e^{-\hbar \omega_0(q)/T} - 1)^{-1}$ ($\omega_0(q) = v_F |q|$)
the initial Bose distribution of modes and where $f$ and $g$ are defined in Eqs. (\ref{eq:fqt}) and (\ref{eq:gqt}).   In the following we shall argue that, in the limit $t \to +\infty$ the above expression
reduces to the following correlator in the generalized Gibbs
ensemble:
\begin{equation}
C_{J_r}^{gG}(x,\tau) = {\rm Tr} \left[ \rho_{\rm gG}(T) \, J_r(x,\tau)J_r(0,0) \right],
\label{eq:ggjr}
\end{equation}
where $\rho_{\rm gG}(T)$ is the extension  to
an initial thermal state of the generalized Gibbs ensemble
 introduced above (notice that since $[H_{\rm f}, I(q)] = 0$ and therefore $[H_{\rm f}, \rho_{\rm gG} ] = 0$,
 it is in principle possible to define time-dependent correlation functions on this ensemble, just
as they are defined  in the equilibrium ensembles). For this (thermal) initial condition
(\ref{eq:inic_cond_T0}) fixes the values of $\lambda(q,T)$, which now
depend on $\beta(q)$ and on the temperature $T$:
\begin{multline}
\sinh^2 \beta(q) \left[1+n_B(q,T)\right]+\cosh^2\beta(q) n_B(q,T)\\=\frac{1}{e^{\lambda(q,T)}-1}.\label{eq:cond}
\end{multline}
Introducing this result into the mode expansion for Eq.~(\ref{eq:ggjr}), we arrive at
\begin{align}
C_{J_r}^{gG}(x,\tau)&=\frac{1}{(2\pi)^2}\sum_{q>0}\left(
\frac{2\pi q}{L}\right) e^{-qa_0}\notag\\
&\times\big\{ e^{iq\left( x-v \tau\right) }\cosh^2\beta(q)\left[
1+\langle n(q)\rangle\right] \notag\\
&+e^{iq\left(x+v\tau\right)  }\sinh^2\beta(q)\ \langle n(q)\rangle\notag \\
&+e^{-iq\left( x-v\tau\right) }\cosh^2\beta(q)\ \langle n(q)\rangle\big\} \notag\\
&+e^{-iq\left( x+v\tau\right)  }\sinh^2\beta(q)\left[ 1+\langle
n(q)\rangle\right].\label{eq:two_points_can_gibbs}
\end{align}
Using the expressions for $f(q,t)$ and $g(q,t)$ given in the Appendix, it can be shown that, in  the limit $t \to +\infty$ of Eq. (\ref{eq:two_points_can}), the rapidly oscillating terms that depend only on $t$ can be dropped, and therefore Eqs. (\ref{eq:two_points_can}) and (\ref{eq:two_points_can_gibbs}) become the same. Thus, the above current
correlation function  relaxes according to the generalized Gibbs ensemble.

 Let us close this section with a digression of higher order
 current correlation functions.  In principle, these correlations
 depend on correlations between $n(q)$ and $n(-q)$  (\emph{i.e.} on $\langle n(q) n(-q)\rangle$),
 which exist in the initial state. These are not captured by the above simple-minded
 generalized Gibbs distribution, which only contains information about the
 expectation value of $n(q)$ in the initial state. However,
 a simple argument based on momentum  conservation and counting powers of $L$
 (the system size) shows that the contribution of these correlations vanishes in the thermodynamic limit, as
 mentioned above.  To illustrate this point, let us consider computing the four point  current correlation function,
 $\langle J_r(x_1, t_1) J_r(x_2, t_2) J_r(x_3, t_3) J_r(x_2,t_4)\rangle_T$.
 Upon using the mode expansion for the current operator $J_r(x,t) = \partial \phi_r(x,t)/2\pi$,
 the correlation function can be expressed as a sum over expectation values of the mode creation
 and destruction operators, the $b$ operators. However, it is important that, since each of these mode expansions
 of the current operator  carries  a factor of  $\frac{1}{\sqrt{L}}$,  the four point correlation function is thus
 proportional to $L^{-2}$ . Upon applying the standard Wick's theorem to the expectation value of the $b$
 operators,  momentum conservation requires that at least two of the four momenta being summed over must be equal.
 In general, the two independent momenta are not equal, and therefore we get a finite contribution in the limit
 $L \to \infty$,  which involves only a product of expectation values of the operator $n(q)$.
 However, when two independent momenta being summed over coincide,  the expectation value becomes
 $\langle n(q) n(-q)\rangle_T$  and thus, in the thermodynamic limit, the sum turns out to be of order $\frac{1}{L}$,
 therefore vanishing as $L\to \infty$. This argument can be extended to higher order correlations of the current operator.
 In a sense, it also justifies the use of Wick's theorem when computing higher order
correlations (in the thermodynamic limit) using the above `simple-minded' generalized Gibbs ensemble.

\section{Relevance to experiments}\label{sec:exp}

 As we described in the introduction, ultracold atom systems
are the ideal arena to study quench dynamics. This is because
they are, to a good approximation,  isolated systems. Furthermore,
as far as one dimensional systems are concerned, there are
already a number of experimental
realizations, including experiments where quench dynamics has been
already studied.\cite{gunter_p_wave_interactions_1D_fermions,stoeferle_shaking%
_fast_tunnability,kinoshita_non_thermalization} Thus, in this section we
would like to  discuss the possible experimental relevance of
the results obtained in previous sections. As mentioned above,
this must be done with great care because our results have been
obtained using a field theory model, namely the Luttinger model (LM),  which
 can be  regarded as `caricature' of the Hamiltonians describing real systems of
ultracold atoms confined to one dimension. We must emphasize that the situation
in the case of quantum quenches in particular,
and of non-equilibrium dynamics in general,  is very different from the analysis
of low-temperature phenomena in equilibrium. In the latter case, the
experimental relevance of models such las the LM  is well established by using
renormalization-group arguments. This  has been put  to test
over the years using a large variety of numerical and also (when possible)
analytical methods. By contrast, in the case of non-equilibrium dynamics
we travel through a largely uncharted land,  and much needs to be understood
in order to achieve a similar level of rigor as in the equilibrium
case.  Thus, it is convenient to regard models as the LM as `toys`,
which can provide us valuable lessons and insights
into non-equilibrium dynamics of strongly correlated systems.
With this cautionary remarks, we can proceed to discuss
some experimental systems for which the above results could be
of some relevance.

 As mentioned in Sect.~\ref{sec:LM} the LM is the exactly solvable model describing
 the renormalization-group fixed point of a general class of interacting
 one-dimensional models\cite{haldane_luttinger_liquid},
known as Tomonaga-Luttinger liquids. This class includes systems such
as the one-dimensional Bose gas interacting via a Dirac-delta potential (which is
solvable via the Bethe-ansatz~\cite{lieb_liniger_model})  as well as many other systems of
interacting  Bose gases with repulsive interactions (such as dipolar) or Fermi gases with both
attractive and repulsive interactions.  With the caveats
of the previous section, it would be interesting to test the results obtained
using the LM in one of these systems. However, the dynamics may be
strongly modified by the fact that higher energy states will be also
excited following a quantum quench. Such higher energy states
are not, in general, correctly described by the LM.  The worse situation  may be expected
in the case of a system where interactions are short ranged,
because, at least from a perturbative point of view, an interaction quench
will couple excited states with the same strength. Thus, one possible way around this problem when trying to
compare with results for the LM model
would be to study  experimentally (or  numerically)  interaction quenches in systems
with long-range  interactions.  This system  provides us with a much
more faithful realization of the LM, which involves long-range interactions.
Furthermore,  the sudden connection of interactions is likely
not to scatter particles to high energy states because interaction decreases with the momentum exchanged
between the colliding particles. One system satisfying this requirement is a single-species dipolar
1D Fermi gas confined to one dimension by a strongly anisotropic trapping potential.  Since for a single species Fermi gas contact interactions in the  $p$-wave channel
are weak away from a $p$-wave Feshbach resonance,
the dominant interaction is the long-ranged dipolar interaction when the dipoles
are all aligned by an external field (electric or magnetic, depending on whether
the dipole is electric, like in hetero-nuclear molecules, or magnetic, like
in Chromium). When confined to  a one-dimensional trap, the dipolar interaction between the atoms
can be approximated by the potential:
\begin{equation}
V_\text{dip}(x,\theta)=\frac{1}{4\pi\epsilon_0}\;
\frac{D^2\lambda(\theta)}{(x^2+R_0^2)^{3/2}}
\end{equation}
where $D$ is the dipolar momentum of the atoms, $\theta$ is the
angle subtended by the direction of the atomic motion and the polarizing
field, and $\lambda(\theta)=(1-3\cos\theta)$. Since in this case
$g_2(q)=g_4(q)\propto\lambda(\theta)$, a sudden change in the
interactions can be produced by a sudden change in alignment of the field
with the direction of motion, that is, a change in $\theta$.  In particular, a  change in $\theta$
away from the value $\theta_m=\cos^{-1}(\frac{1}{3})$ would lead to suddenly switching on
the interactions between the fermions.~\cite{cazalilla_quench_LL}
At zero temperature, the momentum distribution $f(p,t)$  (which can be probed by time of flight measurements)
following the quench into the interacting system would evolve as described in Sect.~\ref{sec:turningon}
(cf. Fig.~\ref{fig:fpt}), with the discontinuity at the Fermi level dying out as $t^{-\gamma^2}$.
However, currently atomic gases  are produced at temperatures $T \sim 10\%$ to $20\%$ of the
Fermi energy, and this would complicate the observation of this effect.
If much lower temperatures could be reached in experiments, so that the application of
the LM becomes  much more reliable,  we expect that in a time of the order of
$\hbar/T$ the quenched dipolar gas reaches a stationary state characterized by a
momentum distribution that differs from the thermal one. However, the calculations
of $f(p,t)$ presented in Sect.~\ref{sec:turningon} (cf. Figs.~\ref{fig:mom_dist_small}
and \ref{fig:mom_dist_large}) show that the differences between the non-equilibrium
and equilibrium results in the stationary state may be well below the current experimental
resolution. Alternatively, instead of measuring the momentum distribution, one can try
to determine the non-equilibrium exponents  by measuring noise correlations in
the time-of-flight images~\cite{mathey_noise_correlations} or through
interferometry.~\cite{polkovnikov_interference_between_condensates}


 To sum up, we have studied correlations in the Luttinger model (LM) following an
interaction quench. By studying the situations where the interactions are suddenly
turned on and off, we have shown that the correlation dynamics is dominated by
the initial state correlations at short times. However, in the thermodynamic limit,
the LM reaches a non-equilibrium steady state characterized by a set of non-equilibrium
exponents that differ from the exponents that describe the decay of the same correlations
in equilibrium. This behavior can be obtained from a `simple-minded'  generalized
Gibbs ensemble.~\cite{JaynesI,JaynesII,rigol_generalized_gibbs_hcbosons,cazalilla_quench_LL}
The correlations also exhibit a `light-cone effect', which was previously observed in correlations following a quantum
quench from a gapped to a critical state.~\cite{calabrese_quench_CFT,calabrese_quench_CFT_long}.
Finally, we have discussed the relevance of our results for the LM for experiments with ultracold
atomic gases confined in highly elongated (one-dimensional) traps. We have argued that the most
appropriate scenario for the observation of the effects described here may be a
one-dimensional dipolar Fermi gas.

\acknowledgments

We thank T. Giamarchi and A. Muramatsu for useful discussions.
AI gratefully acknowledges financial support from
the Swiss National Science Foundation under MaNEP and Division II, CONICET and UNLP
and  hospitality of DIPC, where part of
this work was done. MAC thanks M. Ueda for his kind
hospitality at the University of Tokyo during his visit
at the Ueda  ERATO Macroscopic Quantum Control
Project of JST (Japan),  during which parts of this manuscript
were completed. MAC also gratefully acknowledges financial
support of the  Spanish  MEC through grant No.
FIS2007-66711-C02-02 and CSIC through grant No. PIE 200760/007.

\appendix

\section{Quadratic Hamiltonians}\label{sec:quadratic}

In this appendix we study the case of a
quantum quench in a model described by a quadratic Hamiltonian:
\begin{multline}
H(t) =\sum_{q} \hbar \left[  \omega_0(q) + m(q,t)
\right] b^\dag(q) b(q) \\
\quad + \frac{1}{2} \sum_{q} \hbar g(q,t) \left[ b(q) b( -q) +
b^\dag(q) b^\dag(-q)\right], \label{eq:genham}
\end{multline}
where $[ b(q), b^\dag(q')] = \delta_{q,q'}$, commuting otherwise.
We will assume that the quench takes place at $t = 0$, so that,
within the sudden approximation, the system is described by
$H_\text{i} = H(t \leq 0)$ for $t < 0$ and by $H_\text{f} = H(t >
0)$ for $t  >  0$.  Furthermore, in order to simplify the analysis, we
assume that $m(q,t \leqslant 0) = g(q,t \leqslant 0) = 0$, and
$m(q,t>0) = m(q)$ and $g(q,t) = g(q)$. Notice that the initial
Hamiltonian is diagonal in the $b$ operators:
\begin{equation}
H_\text{i}=H_0\equiv\sum_q
\hbar\omega_0(q)b^\dagger(q)b(q).\label{eq:hamiltonian_H_0}
\end{equation}
In order  to obtain the time evolution of operators  $O =
\mathcal{O}[\{b^\dag(q), b(q) \}]$ after the quench, we recall
that, in the Heisenberg picture, $\mathcal{O}(t > 0) = e^{i
H_\text{f} t/\hbar} \mathcal{O} e^{-i H_\text{f} t/\hbar} =
\mathcal{O}(\{ b(q,t), b^\dag(q,t) \})$, and therefore all that is
needed to solve the above quench problem is to obtain the time
evolution of $b(q)$ for $t > 0$. For Hamiltonians
like~(\ref{eq:genham}) this can be done exactly because
$H_\text{f}  = H(t > 0)$ can be diagonalized by means of the
canonical Bogoliubov (``squeezing'') transformation:
\begin{align}
a(q) & = \cosh \beta(q) \, b(q) + \sinh \beta(q) \, b^\dag(-q).
\label{eq:bogol}
\end{align}
Upon choosing
\begin{equation}
\tanh 2\beta(q) = \frac{g(q)}{\omega_0(q) +
m(q)},\label{eq:bogol_beta}
\end{equation}
the Hamiltonian at $t > 0$ is rendered diagonal:
\begin{equation}
H_\text{f} = H\equiv E_0 +  \sum_{q} \hbar \omega(q) \,
a^{\dag}(q) a(q),\label{eq:ham_diag}
\end{equation}
where $E_0$ is the energy of the ground state of $H$ (relative to
the ground state energy of $H_0$) and
\begin{equation}
\omega(q) = \sqrt{\left[ \omega_0(q) + m(q)\right]^2 - [g(q)]^2}
\end{equation}
the dispersion of the excitations about the ground state of
$H_\text{f}$. The evolution of the $a(q)$ is given by $a(q,t) =
e^{iH_\text{f} t/\hbar} a(q) e^{-iH_\text{f} t/\hbar} = e^{-i
\omega(q) t} a(q)$. By application of a direct and reverse
Bogoliubov transformation, one can obtain the time evolution of
$b(q)$:
\begin{equation}
b(q,t) = f(q,t)\, b(q) + g^*(q,t) \,
b^\dag(-q),\label{eq:solution_time}
\end{equation}
where
\begin{align}
f(q,t) &= \cos \omega(q) t  - i \sin \omega(q) t \, \cosh
2\beta(q),
\label{eq:fqt}\\
g(q,t) &= i \sin \omega(q) t \, \sinh 2\beta(q).\label{eq:gqt}
\end{align}
It is easy to check that (\ref{eq:solution_time}) obeys the
initial condition, $b(q,t =0) = b(q)$, and also respects the
equal-time commutation rules,
\begin{align}
[b(q,t), b(q',t)] &= \big( f(q,t) g^{*}(q,t)
\notag \\
&\quad - g^*(q,t) f(q,t)\big)  \delta_{q,-q} = 0,\\
[b(q,t), b^\dag(q',t) ] &=  \left(  |f(q,t)|^2 - |g(q,t)|^2\right)
\delta_{q,q'} \nonumber \\
&= \delta_{q,q'}
\end{align}
Thus, a quantum quench described by a quadratic Hamiltonian can be
solved by means of a time-dependent canonical transformation.

When the quench is reversed, \emph{i.e.} when the case with $m(q,t
\geqslant 0) = g(q,t \geqslant 0) = 0$, and $m(q,t<0) = m(q)$ and
$g(q,t<0) = g(q)$ is considered, the roles played by the initial
and final Hamiltonians are also reversed: the \emph{final} Hamiltonian
is now diagonal in the $b$'s, $H_\text{f}=H_0$, whereas the
transformation of Eq. (\ref{eq:bogol}) renders diagonal the \emph{initial}
Hamiltonian, $H_\text{i}=H$. Therefore, in this case the
evolution of the $b$ operators is trivial:
$H_\text{f}$: $b(q,t)=e^{-i\omega_0(q)t}$, whereas the evolution of
the $a$'s is given by
\begin{equation}
a(q,t) = f_0(q,t)\, a(q) + g_0^\ast(q,t) \, a^\dag(-q),
\end{equation}
where
\begin{align}
f_0(q,t) &= \cos \omega_0(q) t  - i \sin \omega_0(q) t \, \cosh
2\beta(q),
\label{eq:f0qt}\\
g_0(q,t) &= -i \sin \omega_0(q) t \, \sinh 2\beta(q).\label{eq:g0qt}
\end{align}

\section{Details of the calculation of the one-particle
density matrix in the Luttinger model}\label{app:single_particle_correlations}

In this Appendix, we shall provide  the details of the calculation of non-equilibrium
one-particle density matrix:
\begin{equation}
C_{\psi_r}(x,t) = \langle e^{i H_\text{f} t/\hbar}
\psi^{\dag}_r(x) \psi_r(0) e^{-i H_\text{f} t/\hbar} \rangle,
\end{equation}
To this end, the formula (\ref{eq:bosonization_formula}) is used. In normal ordered form:
\begin{equation}
\psi_\alpha(x) = \frac{e^{- i s_\alpha \pi x/L}}{\sqrt{L}} \,
:e^{i s_\alpha \phi_\alpha(x)}: \quad,
\end{equation}
where the normal order is defined:
\begin{equation}
:e^{i s_\alpha \phi_\alpha(x)}:=e^{i\varphi_\alpha}e^{ 2\pi
i s_\alpha x N_\alpha}e^{ i s_\alpha \Phi_\alpha^\dagger(x)}e^{ i s_\alpha
\Phi_\alpha(x)}.
\end{equation}
The boson field $\Phi_{\alpha}(x)$ is given by  Eq.~(\ref{eq:boson_modes}).
Hence,
\begin{multline}
: e^{-i\phi_r(x)}: \, :e^{+i\phi_r(0)}: \,=  e^{-2\pi i x N_r/L}
e^{\left[ \Phi_r(x), \Phi^{\dag}_r(0)\right]} \\ \times :e^{-i
\left[ \phi_r(x) - \phi_r(0) \right]}:\quad ,
\end{multline}
where we have used the identity $e^{A} e^{B} =  e^{[A,B]}\, e^{B}
e^{A}$, which holds provided $[A,B]$ is a c-number. Using that
($a_0 \to 0^+$ is the short-distance cut-off):
\begin{align}
\left[ \Phi_r(x), \Phi^{\dag}_r(0) \right] &= \sum_{q > 0} \left(
\frac{2\pi}{qL}\right)  e^{-q a_0} e^{i q x}\\
&=  - \ln \left[ 1 - e^{-2\pi a_0/L} e^{2i \pi x/L} \right],
\end{align}
we arrive at  the following expression for $C_{\psi_r}(x,t)$:
\begin{equation}
C_{\psi_r}(x,t) = G^{(0)}_r(x) \, \langle e^{-iF^{\dag}_r(x,t)}
e^{-i F_r(x,t)} \rangle,
\end{equation}
where
\begin{align}
G^{(0)}_r(x) &= \frac{i}{2L} \frac{1}{\sin\left[ \frac{\pi}{L}
\left( x + i a \right)\right]}  \\
F_r(x,t) &= e^{i Ht/\hbar} \left[ \Phi_r(x)   - \Phi_r(0) \right]
e^{-i H_\text{f} t /\hbar} \\
&= \sum_{q > 0} \left( \frac{2\pi}{qL}\right)^{1/2}\,  (e^{i qx} -
1)\\
&\qquad\qquad\times \left[ f(q,t) b(q) + g^\ast(q,t) b^{\dag}(-q)
\right].
\end{align}
To derive the last expression we have used Eq.~(\ref{eq:solution_time}).
Employing the identities $e^{A} e^{B} = e^{[A,B]/2} e^{A+ B}$ (provided
$[A,B]$ is a c-number) and that $\langle e^{D} \rangle =
e^{\langle D\rangle + \frac{1}{2} \langle \left( D - \langle D
\rangle \right)^2\rangle}$, we obtain:
\begin{equation}
C_{\psi_r}(x,t) =  G^{(0)}_r(x) \,  e^{- \langle F^{\dag}_r(x,t)
F_r(x,t) \rangle}.\label{eq:ffi}
\end{equation}
where we have used that $\langle  [F^{\dag}_r(x,t) ]^2 \rangle =
\langle F^2_r(x,t) \rangle =0$ because $ \langle [ b^{\dag}(q) ]^2
\rangle = \langle [ b(q) ]^2 \rangle =  \langle b^{\dag}(-q) b(q)
\rangle = 0$, and since the commutator $[ F^\dagger_r(x,t),
F_r(x,t)]$ is a c-number, it can be safely replaced by $\langle [
F^\dagger_r(x,t), F_r(x,t) ] \rangle$. Note that  for $t = 0$
$F_r(x,t =0)$ contains only $b(q)$ and thus the average $\langle
F^{\dag}_r(x,t) F_r(x,t) \rangle = 0$ at $ T= 0$. In Eq.~(\ref{eq:ffi})
the exponent can be expanded to yield:
\begin{multline}
\langle F^{\dag}_r(x,t) F_r(x,t) \rangle = \sum_{q > 0} \left(
\frac{2\pi}{qL}\right) e^{-q a_0} | e^{iqx} - 1|^2 \\
\times\left[ |f(q,t)|^2 n_B(q)  + |g(q,t)|^2  \left( n_B(q)
+1\right) \right],\label{eq:ff}
\end{multline}
being $n_B(q) = \langle a^{\dag}(q) a(q) \rangle = (e^{-\lambda |q|} - 1)^{-1}$ the
distribution of Tomonaga bosons in the initial state
(which has been assumed to be a mixed thermal  state),
and  $\lambda=\hbar v_F/T$ is the thermal correlation length.
We next evaluate explicitly the above result in
several limiting cases.

\subsection{Zero temperature and finite length}

Let us now consider the $T = 0$ limit of the above correlator, where $n_B(q) = 0$, and
thus, using (\ref{eq:fqt}) and (\ref{eq:gqt}), Eq. (\ref{eq:ff}) simplifies to:
\begin{multline}\label{eq:func_F}
\langle F^{\dag}_r(x,t) F_r(x,t) \rangle_{T=0} = \sum_{q > 0}
\left( \frac{2\pi}{qL}\right) e^{-qa_0}\sinh^2[2\beta(q)] \\
\times  \left( 1 - \cos q x \right) [1-\cos 2 v(q) |q| t].
\end{multline}
To make further progress, we  assume that $\sinh 2 \beta(q) =
\gamma e^{-|q R_0|/2} $ where $R_0$ is the range of the
interaction. Furthermore, we replace $v(q)$ by $v =
v(0)$~\footnote{We assume implicitly that both $\beta(q)$ and
$v(q)$ are not singular at $q = 0$.}, what allows us to safely
take the limit $a _0\to 0+$. Next, in order to simplify the
computation, we introduce the quantity
\begin{equation}
{\cal E}_r(z)=\sum_{q > 0} \left( \frac{2\pi}{qL} \right) e^{-q R_0}
\cos qz,
\end{equation}
which can be readily computed to give
\begin{equation}
{\cal E}_r(z)=-\ln\left[\frac{\pi}{L}d(z+i R_0|L)\right]+\frac{\pi
R_0}{L}-\ln2,
\end{equation}
where $d(z|L)=L|\sin(\pi z/L)|/\pi$ is the \emph{cord} function.
Using this result into Eq. (\ref{eq:func_F}), yields the following
expression for the one-particle density matrix:
\begin{multline}\label{eq:final_l}
C_{\psi_r}(x,t>0|L)=G_r^{(0)}(x|L)\left\vert\frac{d(i R_0|L)}
{d(x+i R_0|L)}\right\vert^{\gamma^2}\\
\times\left\vert\frac{d(x+2vt+i R_0|L)d(x-2vt+i R_0|L)}{d(2vt+i
R_0|L)d(-2vt+i R_0|L)}\right\vert^{\gamma^2/2}.
\end{multline}
Taking into account that $R_0/L \ll 1$ and,
we obtain the result quoted  in the main text,
Eq.~(\ref{eq:green_function_LM}), in the scaling limit.

\subsection{Thermodynamic limit and finite temperature}

We next consider Eq.(\ref{eq:ffi}) for $L \to \infty$ and finite temperature, $T$.
Equation (\ref{eq:ff}) can be recast as:
\begin{multline}\label{eq:func_F_finite_T}
\langle F^{\dag}_r(x,t) F_r(x,t) \rangle_T = \langle F_r(x,t)
F_r(x,t) \rangle_{T = 0}\\ + {\cal H}(x) + {\cal G}(x,t)
\end{multline}
where we have introduced the following functions:

\begin{align}
{\cal H}(x) &= 2 \int^{\infty}_0 \frac{dq}{q} e^{-q a}\left(1-\cos qx \right)n_B(q), \\
{\cal G}(x,t) &= 2\gamma^2 \int^{\infty}_0 \frac{dq}{q} e^{-q R_0}
\left( 1 - \cos qx \right)\\ &\qquad\qquad\qquad [1-\cos (2 v q
t)] \, n_B(q),\label{eq:func_G}
\end{align}
which hold in the thermodynamic limit and upon replacing $v(q)$ by
$v = v(q =0)$ and $\sinh 2\beta(q) = \gamma e^{-|q R_0|/2}$ as we did
in the previous section. We next define the function
\begin{equation}
g(u;r) = 2 \int^{+\infty}_0 \frac{dq}{q} e^{-qr} \, \frac{(1 -
\cos qu)}{e^{\lambda p}-1}.
\end{equation}
which can be evaluated to yield: \cite{gradshteyn80_tables}
\begin{equation}
g(u;r) =
2\ln\left|\frac{\Gamma(1+\lambda^{-1}r)}{\Gamma[1+\lambda^{-1}(r+iu)]}\right|,
\end{equation}
where $\Gamma(z)$ is the Gamma function. In the limit where $r\ll u$,
and using that\cite{gradshteyn80_tables}
$\Gamma(z)\Gamma(1-z)=\pi/\sin(\pi z)$, the above expression
reduces to
\begin{equation}\label{eq:func_h}
g(u;r) = -\ln\left\vert\frac{dh(ir|T)}{dh(u+ir|T)}\right\vert
-\ln\left\vert\frac{u+ir}{r}\right\vert.
\end{equation}
In the previous expression we have defined:
\begin{equation}
dh(z|T)=\frac{\lambda}{\pi}\left\vert\sinh\left(\pi\lambda^{-1}z\right)\right\vert.
\end{equation}
Combining this result with Eqs.~(\ref{eq:func_F_finite_T})-(\ref{eq:func_G}) and (\ref{eq:ffi}),
it is seen that the second term in Eq.~(\ref{eq:func_h}) exactly cancels
the contributions from $G^{(0)}(x|L)$ and $\langle F_r(x,t)
F_r(x,t) \rangle_{T = 0}$ in the thermodynamic limit, and
therefore,
\begin{multline}
C_{\psi_r}(x,t|T) =
G^{(0)}_r(x|T)\left[\frac{dh(i R_0|T)}{dh(x+i R_0|T)}\right]^{\gamma^2}\\
\times\left[\frac{dh(x+2vt+i R_0|T)dh(x-2vt+i R_0|T)}{dh(2vt+i
R_0|T)dh(-2vt+i R_0|T)}\right]^{\gamma^2/2}. \label{eq:cpsiT}
\end{multline}
where
\begin{equation}
G^{(0)}_r(x|T) = \frac{i}{2\pi} \frac{\pi
\lambda^{-1}}{\sinh\left[ \pi\lambda^{-1}\left( x + i a_0
\right)\right]}.
\end{equation}
We note that the result of Eq.~(\ref{eq:cpsiT}) can be obtained from Eq.
(\ref{eq:final_l}) upon making the replacement $L\sin(\pi L^{-1}x)/\pi$ by $
\lambda\sinh(\pi \lambda^{-1} x)$. Taking into account that
$R_0/L \ll 1$,  we retrieve the result quoted
in the main text, Eq.~(\ref{eq:corr_func_fin_T}), in the scaling limit.

\section{One-body density matrix of the Luttinger model
in the generalized Gibbs ensemble}\label{ap:green_function_gibbs}

 Next we take up the calculation of the one-body density matrix in the
generalized Gibbs ensemble for the Luttinger model discussed in
Sect.~\ref{sec:generalized}. That is, we shall evaluate the expression
at $T = 0$.
\begin{equation}
C^{\rm gG}_{\psi_r}(x) =  \Tr\left[ \rho_{\rm gG} \:  \psi^{\dag}_r(x) \psi_r(0)\right]
\end{equation}
Using the bosonization identity, Eq.~(\ref{eq:bosonization_formula}),
we can write the expression as follows:
\begin{align}
C^{\rm gG}_{\psi_r}(x) &=  G^{(0)}_r(x) \: \langle :e^{-i\left[ \phi_r(x)
- \phi_r(0) \right]}: \rangle_\text{gG} \\
&= G^{(0)}_r(x)\, \langle e^{-i\tilde{F}^{\dag}_r(x)} e^{-i
\tilde{F}_r(x)} \rangle_{\rm gG}.
\end{align}
Taking into account that
\begin{align}
\tilde{F}_r(x) &= \Phi_r(x) - \Phi_r(0) \\
&=  \sum_{q > 0} \left( \frac{2\pi}{qL}\right)^{1/2} e^{-q a/2}
(e^{iq x} - 1) \\
&\quad\quad\times \left[ \cosh \beta(q) a(q)  - \sinh \beta(q)
a^{\dag}(-q) \right].
\end{align}
The expression for $C^{\rm gG}_{\psi_r}(x)$ can be easily computed by
using the trick of regarding $\rho_{gG}$ as a canonical ensemble with
$q$-dependent temperature. Thus, following
the same steps as in the previous section we arrive at:
\begin{equation}
C^{\rm gG}_{\psi_r}(x) = G^{(0)}_r(x) \: e^{-\langle \tilde{F}^{\dag}_r(x)
\tilde{F}_r(x) \rangle_\text{gG}}.
\end{equation}
Given that
\begin{equation}
\langle \tilde{F}^{\dag}_r(x) \tilde{F}_r(x) \rangle_\text{gG}=
\sinh^2 2\beta \left[ \mathcal{D}_{r}(0) - \mathcal{D}_r(x) \right],
\end{equation}
where
\begin{align}
\mathcal{D}_r(x) &= \Re\left\{\sum_{q>0}\left(\frac{2\pi}{qL}\right)e^{-q R_0}e^{iqx}\right\}\\
&= - \ln \left| \sin \frac{\pi}{L} (x + i R_0) \right| - \ln 2 -
\frac{\pi R_0}{L}.
\end{align}
Hence, taking the thermodynamic limit
\begin{equation}
C^{\rm gG}_{\psi_r}(x)  = \frac{i}{2\pi (x + i a)} \left| \frac{R_0}{x}\right|^{\gamma^2}.
\end{equation}
Thus we see that one recovers the same results as $\lim_{t \to +\infty} C_{\psi_r}(x,t)$,
Eq.~(\ref{eq:cpsi}).


\begin{thebibliography}{64}
\expandafter\ifx\csname natexlab\endcsname\relax\def\natexlab#1{#1}\fi
\expandafter\ifx\csname bibnamefont\endcsname\relax
  \def\bibnamefont#1{#1}\fi
\expandafter\ifx\csname bibfnamefont\endcsname\relax
  \def\bibfnamefont#1{#1}\fi
\expandafter\ifx\csname citenamefont\endcsname\relax
  \def\citenamefont#1{#1}\fi
\expandafter\ifx\csname url\endcsname\relax
  \def\url#1{\texttt{#1}}\fi
\expandafter\ifx\csname urlprefix\endcsname\relax\def\urlprefix{URL }\fi
\providecommand{\bibinfo}[2]{#2}
\providecommand{\eprint}[2][]{\url{#2}}

\bibitem[{\citenamefont{Altman and Auerbach}(2002)}]{altman_quench_2002}
\bibinfo{author}{\bibfnamefont{E.}~\bibnamefont{Altman}} \bibnamefont{and}
  \bibinfo{author}{\bibfnamefont{A.}~\bibnamefont{Auerbach}},
  \bibinfo{journal}{Phys. Rev. Lett.} \textbf{\bibinfo{volume}{89}},
  \bibinfo{pages}{250404} (\bibinfo{year}{2002}).

\bibitem[{\citenamefont{Sengupta et~al.}(2004)\citenamefont{Sengupta, Powell,
  and Sachdev}}]{sengupta_quench_QCP}
\bibinfo{author}{\bibfnamefont{K.}~\bibnamefont{Sengupta}},
  \bibinfo{author}{\bibfnamefont{S.}~\bibnamefont{Powell}}, \bibnamefont{and}
  \bibinfo{author}{\bibfnamefont{S.}~\bibnamefont{Sachdev}},
  \bibinfo{journal}{Phys. Rev. A} \textbf{\bibinfo{volume}{69}},
  \bibinfo{pages}{053616} (\bibinfo{year}{2004}).

\bibitem[{\citenamefont{Barankov and
  Levitov}(2006)}]{barankov_dynamical_projection}
\bibinfo{author}{\bibfnamefont{R.~A.} \bibnamefont{Barankov}} \bibnamefont{and}
  \bibinfo{author}{\bibfnamefont{L.~S.} \bibnamefont{Levitov}},
  \bibinfo{journal}{Phys. Rev. Lett.} \textbf{\bibinfo{volume}{96}},
  \bibinfo{pages}{230403} (\bibinfo{year}{2006}).

\bibitem[{\citenamefont{Yuzbashyan et~al.}(2005)\citenamefont{Yuzbashyan,
  Altshuler, Kuznetsov, and Enolskii}}]{yuzbashyan_BCS_quench}
\bibinfo{author}{\bibfnamefont{E.~A.} \bibnamefont{Yuzbashyan}},
  \bibinfo{author}{\bibfnamefont{B.~L.} \bibnamefont{Altshuler}},
  \bibinfo{author}{\bibfnamefont{V.~B.} \bibnamefont{Kuznetsov}},
  \bibnamefont{and} \bibinfo{author}{\bibfnamefont{V.~Z.}
  \bibnamefont{Enolskii}}, \bibinfo{journal}{Phys. Rev. B}
  \textbf{\bibinfo{volume}{72}}, \bibinfo{pages}{220503(R)}
  (\bibinfo{year}{2005}).

\bibitem[{\citenamefont{Kollath et~al.}(2005)\citenamefont{Kollath,
  Schollw{\"o}ck, von Delft, and Zwerger}}]{kollath_density_waves}
\bibinfo{author}{\bibfnamefont{C.}~\bibnamefont{Kollath}},
  \bibinfo{author}{\bibfnamefont{U.}~\bibnamefont{Schollw{\"o}ck}},
  \bibinfo{author}{\bibfnamefont{J.}~\bibnamefont{von Delft}},
  \bibnamefont{and} \bibinfo{author}{\bibfnamefont{W.}~\bibnamefont{Zwerger}},
  \bibinfo{journal}{Phys. Rev. A} \textbf{\bibinfo{volume}{71}},
  \bibinfo{pages}{053606} (\bibinfo{year}{2005}).

\bibitem[{\citenamefont{Altman and
  Vishwanath}(2005)}]{altman_projection_feshbach_molecules}
\bibinfo{author}{\bibfnamefont{E.}~\bibnamefont{Altman}} \bibnamefont{and}
  \bibinfo{author}{\bibfnamefont{A.}~\bibnamefont{Vishwanath}},
  \bibinfo{journal}{Phys. Rev. Lett.} \textbf{\bibinfo{volume}{95}},
  \bibinfo{pages}{110404} (\bibinfo{year}{2005}).

\bibitem[{\citenamefont{Ruschhaupt et~al.}(2006)\citenamefont{Ruschhaupt, {del
  Campo}, and Muga}}]{ruschhaupt_quench_momentum_interference}
\bibinfo{author}{\bibfnamefont{A.}~\bibnamefont{Ruschhaupt}},
  \bibinfo{author}{\bibfnamefont{A.}~\bibnamefont{{del Campo}}},
  \bibnamefont{and} \bibinfo{author}{\bibfnamefont{J.~G.} \bibnamefont{Muga}},
  \bibinfo{journal}{Eur. Phys. J. D} \textbf{\bibinfo{volume}{40}},
  \bibinfo{pages}{399} (\bibinfo{year}{2006}).

\bibitem[{\citenamefont{Yuzbashyan and
  Dzero}(2006)}]{yuzbashyan_quench_fermions_order_parameter}
\bibinfo{author}{\bibfnamefont{E.~A.} \bibnamefont{Yuzbashyan}}
  \bibnamefont{and} \bibinfo{author}{\bibfnamefont{M.}~\bibnamefont{Dzero}},
  \bibinfo{journal}{Phys. Rev. Lett.} \textbf{\bibinfo{volume}{96}},
  \bibinfo{pages}{230404} (\bibinfo{year}{2006}).

\bibitem[{\citenamefont{Cazalilla}(2006)}]{cazalilla_quench_LL}
\bibinfo{author}{\bibfnamefont{M.~A.} \bibnamefont{Cazalilla}},
  \bibinfo{journal}{Phys. Rev. Lett.} \textbf{\bibinfo{volume}{97}},
  \bibinfo{pages}{156403} (\bibinfo{year}{2006}).

\bibitem[{\citenamefont{Rigol et~al.}(2007{\natexlab{a}})\citenamefont{Rigol,
  Dunjko, Yurovsky, and Olshanii}}]{rigol_generalized_gibbs_hcbosons}
\bibinfo{author}{\bibfnamefont{M.}~\bibnamefont{Rigol}},
  \bibinfo{author}{\bibfnamefont{V.}~\bibnamefont{Dunjko}},
  \bibinfo{author}{\bibfnamefont{V.}~\bibnamefont{Yurovsky}}, \bibnamefont{and}
  \bibinfo{author}{\bibfnamefont{M.}~\bibnamefont{Olshanii}},
  \bibinfo{journal}{Phys. Rev. Lett.} \textbf{\bibinfo{volume}{98}},
  \bibinfo{pages}{050405} (\bibinfo{year}{2007}{\natexlab{a}}).

\bibitem[{\citenamefont{Perfetto}(2006)}]{perfetto_quench_coupled_LL}
\bibinfo{author}{\bibfnamefont{E.}~\bibnamefont{Perfetto}},
  \bibinfo{journal}{Phys. Rev. B} \textbf{\bibinfo{volume}{74}},
  \bibinfo{pages}{205123} (\bibinfo{year}{2006}).

\bibitem[{\citenamefont{Calabrese and Cardy}(2006)}]{calabrese_quench_CFT}
\bibinfo{author}{\bibfnamefont{P.}~\bibnamefont{Calabrese}} \bibnamefont{and}
  \bibinfo{author}{\bibfnamefont{J.}~\bibnamefont{Cardy}},
  \bibinfo{journal}{Phys. Rev. Lett.} \textbf{\bibinfo{volume}{96}},
  \bibinfo{pages}{136801} (\bibinfo{year}{2006}).

\bibitem[{\citenamefont{Rigol et~al.}(2006)\citenamefont{Rigol, Muramatsu, and
  Olshanii}}]{Rigol_hc_noneq2}
\bibinfo{author}{\bibfnamefont{M.}~\bibnamefont{Rigol}},
  \bibinfo{author}{\bibfnamefont{A.}~\bibnamefont{Muramatsu}},
  \bibnamefont{and} \bibinfo{author}{\bibfnamefont{M.}~\bibnamefont{Olshanii}},
  \bibinfo{journal}{Phys. Rev. A} \textbf{\bibinfo{volume}{74}},
  \bibinfo{pages}{053616} (\bibinfo{year}{2006}).

\bibitem[{\citenamefont{Manmana et~al.}(2007)\citenamefont{Manmana, Wessel,
  Noack, and Muramatsu}}]{manmana_quench_spinless_fermions_nnn}
\bibinfo{author}{\bibfnamefont{S.~R.} \bibnamefont{Manmana}},
  \bibinfo{author}{\bibfnamefont{S.}~\bibnamefont{Wessel}},
  \bibinfo{author}{\bibfnamefont{R.~M.} \bibnamefont{Noack}}, \bibnamefont{and}
  \bibinfo{author}{\bibfnamefont{A.}~\bibnamefont{Muramatsu}},
  \bibinfo{journal}{Phys. Rev. Lett.} \textbf{\bibinfo{volume}{98}},
  \bibinfo{pages}{210405} (\bibinfo{year}{2007}).

\bibitem[{\citenamefont{Calabrese and Cardy}(2007)}]{calabrese_quench_CFT_long}
\bibinfo{author}{\bibfnamefont{P.}~\bibnamefont{Calabrese}} \bibnamefont{and}
  \bibinfo{author}{\bibfnamefont{J.}~\bibnamefont{Cardy}}, \bibinfo{journal}{J.
  Stat. Mech.: Theor. Exp.} p. \bibinfo{pages}{P06008} (\bibinfo{year}{2007}).

\bibitem[{\citenamefont{Kollath et~al.}(2007)\citenamefont{Kollath,
  L{\"a}uchli, and Altman}}]{kollath_quench_BH}
\bibinfo{author}{\bibfnamefont{C.}~\bibnamefont{Kollath}},
  \bibinfo{author}{\bibfnamefont{A.~M.} \bibnamefont{L{\"a}uchli}},
  \bibnamefont{and} \bibinfo{author}{\bibfnamefont{E.}~\bibnamefont{Altman}},
  \bibinfo{journal}{Phys. Rev. Lett.} \textbf{\bibinfo{volume}{98}},
  \bibinfo{pages}{180601} (\bibinfo{year}{2007}).

\bibitem[{\citenamefont{Gritsev et~al.}(2007)\citenamefont{Gritsev, Demler,
  Lukin, and Polkovnikov}}]{Gritsev_spectroscopy_07}
\bibinfo{author}{\bibfnamefont{V.}~\bibnamefont{Gritsev}},
  \bibinfo{author}{\bibfnamefont{E.}~\bibnamefont{Demler}},
  \bibinfo{author}{\bibfnamefont{M.}~\bibnamefont{Lukin}}, \bibnamefont{and}
  \bibinfo{author}{\bibfnamefont{A.}~\bibnamefont{Polkovnikov}},
  \bibinfo{journal}{Phys. Rev. Lett.} \textbf{\bibinfo{volume}{99}},
  \bibinfo{pages}{200404} (\bibinfo{year}{2007}).

\bibitem[{\citenamefont{Eckstein and
  Kollar}(2008)}]{eckstein_generalized_gibbs_FK}
\bibinfo{author}{\bibfnamefont{M.}~\bibnamefont{Eckstein}} \bibnamefont{and}
  \bibinfo{author}{\bibfnamefont{M.}~\bibnamefont{Kollar}},
  \bibinfo{journal}{Phys. Rev. Lett.} \textbf{\bibinfo{volume}{100}},
  \bibinfo{pages}{120404} (\bibinfo{year}{2008}).

\bibitem[{\citenamefont{Kollar and
  Eckstein}(2008)}]{eckstein_generalized_gibbs_hubbard}
\bibinfo{author}{\bibfnamefont{M.}~\bibnamefont{Kollar}} \bibnamefont{and}
  \bibinfo{author}{\bibfnamefont{M.}~\bibnamefont{Eckstein}},
  \bibinfo{journal}{Phys. Rev. A} \textbf{\bibinfo{volume}{78}},
  \bibinfo{pages}{013626} (\bibinfo{year}{2008}).

\bibitem[{\citenamefont{Moeckel and Kehrein}(2008)}]{kehrein_quench_hubbard}
\bibinfo{author}{\bibfnamefont{M.}~\bibnamefont{Moeckel}} \bibnamefont{and}
  \bibinfo{author}{\bibfnamefont{S.}~\bibnamefont{Kehrein}},
  \bibinfo{journal}{Phys. Rev. Lett.} \textbf{\bibinfo{volume}{100}},
  \bibinfo{pages}{175702} (\bibinfo{year}{2008}).

\bibitem[{\citenamefont{Cramer et~al.}(2008)\citenamefont{Cramer, Flesch,
  McCulloch, Schollw\"ock, and Eisert}}]{eisert_crap1}
\bibinfo{author}{\bibfnamefont{M.}~\bibnamefont{Cramer}},
  \bibinfo{author}{\bibfnamefont{A.}~\bibnamefont{Flesch}},
  \bibinfo{author}{\bibfnamefont{I.~P.} \bibnamefont{McCulloch}},
  \bibinfo{author}{\bibfnamefont{U.}~\bibnamefont{Schollw\"ock}},
  \bibnamefont{and} \bibinfo{author}{\bibfnamefont{J.}~\bibnamefont{Eisert}},
  \bibinfo{journal}{Phys. Rev. Lett.} \textbf{\bibinfo{volume}{101}},
  \bibinfo{pages}{063001} (\bibinfo{year}{2008}).

\bibitem[{\citenamefont{Flesch et~al.}(2008)\citenamefont{Flesch, Cramer,
  McCulloch, Schollw{\"o}ck, and Eisert}}]{eisert_crap2}
\bibinfo{author}{\bibfnamefont{A.}~\bibnamefont{Flesch}},
  \bibinfo{author}{\bibfnamefont{M.}~\bibnamefont{Cramer}},
  \bibinfo{author}{\bibfnamefont{I.~P.} \bibnamefont{McCulloch}},
  \bibinfo{author}{\bibfnamefont{U.}~\bibnamefont{Schollw{\"o}ck}},
  \bibnamefont{and} \bibinfo{author}{\bibfnamefont{J.}~\bibnamefont{Eisert}},
  \bibinfo{journal}{Phys. Rev. A} \textbf{\bibinfo{volume}{78}},
  \bibinfo{pages}{033608} (\bibinfo{year}{2008}).

\bibitem[{\citenamefont{Rigol et~al.}(2007{\natexlab{b}})\citenamefont{Rigol,
  Dunjko, Yurovsky, and Olshanii}}]{rigol_noninteg}
\bibinfo{author}{\bibfnamefont{M.}~\bibnamefont{Rigol}},
  \bibinfo{author}{\bibfnamefont{V.}~\bibnamefont{Dunjko}},
  \bibinfo{author}{\bibfnamefont{V.}~\bibnamefont{Yurovsky}}, \bibnamefont{and}
  \bibinfo{author}{\bibfnamefont{M.}~\bibnamefont{Olshanii}},
  \bibinfo{journal}{Nature (London)} \textbf{\bibinfo{volume}{452}},
  \bibinfo{pages}{854} (\bibinfo{year}{2007}{\natexlab{b}}).

\bibitem[{\citenamefont{Manmana et~al.}(2009)\citenamefont{Manmana, Wessel,
  Noack, and Muramatsu}}]{Manmana_long}
\bibinfo{author}{\bibfnamefont{S.~R.} \bibnamefont{Manmana}},
  \bibinfo{author}{\bibfnamefont{S.}~\bibnamefont{Wessel}},
  \bibinfo{author}{\bibfnamefont{R.~M.} \bibnamefont{Noack}}, \bibnamefont{and}
  \bibinfo{author}{\bibfnamefont{A.}~\bibnamefont{Muramatsu}},
  \bibinfo{journal}{Phys. Rev. B} \textbf{\bibinfo{volume}{79}},
  \bibinfo{pages}{155104} (\bibinfo{year}{2009}).

\bibitem[{\citenamefont{{De Grandi} et~al.}(2008)\citenamefont{{De Grandi},
  Barankov, and Polkovnikov}}]{degrandi_adiabatic_quench}
\bibinfo{author}{\bibfnamefont{C.}~\bibnamefont{{De Grandi}}},
  \bibinfo{author}{\bibfnamefont{R.~A.} \bibnamefont{Barankov}},
  \bibnamefont{and}
  \bibinfo{author}{\bibfnamefont{A.}~\bibnamefont{Polkovnikov}},
  \bibinfo{journal}{Phys. Rev. Lett.} \textbf{\bibinfo{volume}{101}},
  \bibinfo{pages}{230402} (\bibinfo{year}{2008}).

\bibitem[{\citenamefont{Reimann}(2008)}]{Reimann08}
\bibinfo{author}{\bibfnamefont{P.}~\bibnamefont{Reimann}},
  \bibinfo{journal}{Phys. Rev. Lett.} \textbf{\bibinfo{volume}{101}},
  \bibinfo{pages}{190403} (\bibinfo{year}{2008}).

\bibitem[{\citenamefont{Faribault et~al.}(2009)\citenamefont{Faribault,
  Calabrese, and Caux}}]{faribault_fermion_pairing_model}
\bibinfo{author}{\bibfnamefont{A.}~\bibnamefont{Faribault}},
  \bibinfo{author}{\bibfnamefont{P.}~\bibnamefont{Calabrese}},
  \bibnamefont{and} \bibinfo{author}{\bibfnamefont{J.-S.} \bibnamefont{Caux}},
  \bibinfo{journal}{J. Stat. Mech.: Theor. Exp.} p. \bibinfo{pages}{P03018}
  (\bibinfo{year}{2009}).

\bibitem[{\citenamefont{Patan\`e
  et~al.}(2009{\natexlab{a}})\citenamefont{Patan\`e, Amico, Silva, Fazio, and
  Santoro}}]{Patane08}
\bibinfo{author}{\bibfnamefont{D.}~\bibnamefont{Patan\`e}},
  \bibinfo{author}{\bibfnamefont{L.}~\bibnamefont{Amico}},
  \bibinfo{author}{\bibfnamefont{A.}~\bibnamefont{Silva}},
  \bibinfo{author}{\bibfnamefont{R.}~\bibnamefont{Fazio}}, \bibnamefont{and}
  \bibinfo{author}{\bibfnamefont{G.~E.} \bibnamefont{Santoro}},
  \bibinfo{journal}{Phys. Rev. B}  (\bibinfo{year}{2009}{\natexlab{a}}).

\bibitem[{\citenamefont{Patan\`e
  et~al.}(2009{\natexlab{b}})\citenamefont{Patan\`e, Silva, Sols, and
  Amico}}]{Patane08b}
\bibinfo{author}{\bibfnamefont{D.}~\bibnamefont{Patan\`e}},
  \bibinfo{author}{\bibfnamefont{A.}~\bibnamefont{Silva}},
  \bibinfo{author}{\bibfnamefont{F.}~\bibnamefont{Sols}}, \bibnamefont{and}
  \bibinfo{author}{\bibfnamefont{L.}~\bibnamefont{Amico}},
  \bibinfo{journal}{Phys. Rev. Lett.} \textbf{\bibinfo{volume}{102}},
  \bibinfo{pages}{245701} (\bibinfo{year}{2009}{\natexlab{b}}).

\bibitem[{\citenamefont{Rossini et~al.}(2009)\citenamefont{Rossini, Silva, ,
  Mussardo, and Santoro}}]{Rossini08}
\bibinfo{author}{\bibfnamefont{D.}~\bibnamefont{Rossini}},
  \bibinfo{author}{\bibfnamefont{A.}~\bibnamefont{Silva}}, ,
  \bibinfo{author}{\bibfnamefont{G.}~\bibnamefont{Mussardo}}, \bibnamefont{and}
  \bibinfo{author}{\bibfnamefont{G.}~\bibnamefont{Santoro}},
  \bibinfo{journal}{Phys. Rev. Lett.} \textbf{\bibinfo{volume}{102}},
  \bibinfo{pages}{127204} (\bibinfo{year}{2009}).

\bibitem[{\citenamefont{Barmettler et~al.}(2009)\citenamefont{Barmettler, Punk,
  Gritsev, Demler, and Altman}}]{Barmettler08}
\bibinfo{author}{\bibfnamefont{P.}~\bibnamefont{Barmettler}},
  \bibinfo{author}{\bibfnamefont{M.}~\bibnamefont{Punk}},
  \bibinfo{author}{\bibfnamefont{V.}~\bibnamefont{Gritsev}},
  \bibinfo{author}{\bibfnamefont{E.}~\bibnamefont{Demler}}, \bibnamefont{and}
  \bibinfo{author}{\bibfnamefont{E.}~\bibnamefont{Altman}},
  \bibinfo{journal}{Phys. Rev. Lett.} \textbf{\bibinfo{volume}{102}},
  \bibinfo{pages}{130603} (\bibinfo{year}{2009}).

\bibitem[{\citenamefont{Sen et~al.}(2008)\citenamefont{Sen, Sengupta, and
  Mondal}}]{sen_nonlinear_quench_QCP}
\bibinfo{author}{\bibfnamefont{D.}~\bibnamefont{Sen}},
  \bibinfo{author}{\bibfnamefont{K.}~\bibnamefont{Sengupta}}, \bibnamefont{and}
  \bibinfo{author}{\bibfnamefont{S.}~\bibnamefont{Mondal}},
  \bibinfo{journal}{Phys. Rev. Lett.} \textbf{\bibinfo{volume}{101}},
  \bibinfo{pages}{016806} (\bibinfo{year}{2008}).

\bibitem[{\citenamefont{Divakaran et~al.}(2008)\citenamefont{Divakaran, Dutta,
  and Sen}}]{sen_2}
\bibinfo{author}{\bibfnamefont{U.}~\bibnamefont{Divakaran}},
  \bibinfo{author}{\bibfnamefont{A.}~\bibnamefont{Dutta}}, \bibnamefont{and}
  \bibinfo{author}{\bibfnamefont{D.}~\bibnamefont{Sen}},
  \bibinfo{journal}{Phys. Rev. B} \textbf{\bibinfo{volume}{78}},
  \bibinfo{pages}{144301} (\bibinfo{year}{2008}).

\bibitem[{\citenamefont{Mondal et~al.}(2009)\citenamefont{Mondal, Sengupta, and
  Sen}}]{sen_3}
\bibinfo{author}{\bibfnamefont{S.}~\bibnamefont{Mondal}},
  \bibinfo{author}{\bibfnamefont{K.}~\bibnamefont{Sengupta}}, \bibnamefont{and}
  \bibinfo{author}{\bibfnamefont{D.}~\bibnamefont{Sen}},
  \bibinfo{journal}{Phys. Rev. B} \textbf{\bibinfo{volume}{79}},
  \bibinfo{pages}{045128} (\bibinfo{year}{2009}).

\bibitem[{\citenamefont{Divakaran et~al.}(2009)\citenamefont{Divakaran,
  Mukherjee, Dutta, and Sen}}]{sen_4}
\bibinfo{author}{\bibfnamefont{U.}~\bibnamefont{Divakaran}},
  \bibinfo{author}{\bibfnamefont{V.}~\bibnamefont{Mukherjee}},
  \bibinfo{author}{\bibfnamefont{A.}~\bibnamefont{Dutta}}, \bibnamefont{and}
  \bibinfo{author}{\bibfnamefont{D.}~\bibnamefont{Sen}}, \bibinfo{journal}{J.
  Stat. Mech.}  (\bibinfo{year}{2009}).

\bibitem[{\citenamefont{Sengupta et~al.}(2008)\citenamefont{Sengupta, Sen, and
  Mondal}}]{sen_5}
\bibinfo{author}{\bibfnamefont{K.}~\bibnamefont{Sengupta}},
  \bibinfo{author}{\bibfnamefont{D.}~\bibnamefont{Sen}}, \bibnamefont{and}
  \bibinfo{author}{\bibfnamefont{S.}~\bibnamefont{Mondal}},
  \bibinfo{journal}{Phys. Rev. Lett.} \textbf{\bibinfo{volume}{100}},
  \bibinfo{pages}{077204} (\bibinfo{year}{2008}).

\bibitem[{\citenamefont{Mukherjee et~al.}(2007)\citenamefont{Mukherjee,
  Divakaran, Dutta, and Sen}}]{sen_6}
\bibinfo{author}{\bibfnamefont{V.}~\bibnamefont{Mukherjee}},
  \bibinfo{author}{\bibfnamefont{U.}~\bibnamefont{Divakaran}},
  \bibinfo{author}{\bibfnamefont{A.}~\bibnamefont{Dutta}}, \bibnamefont{and}
  \bibinfo{author}{\bibfnamefont{D.}~\bibnamefont{Sen}},
  \bibinfo{journal}{Phys. Rev. B} \textbf{\bibinfo{volume}{76}},
  \bibinfo{pages}{174303} (\bibinfo{year}{2007}).

\bibitem[{\citenamefont{Mukherjee et~al.}(2008)\citenamefont{Mukherjee, Dutta,
  and Sen}}]{sen_7}
\bibinfo{author}{\bibfnamefont{V.}~\bibnamefont{Mukherjee}},
  \bibinfo{author}{\bibfnamefont{A.}~\bibnamefont{Dutta}}, \bibnamefont{and}
  \bibinfo{author}{\bibfnamefont{D.}~\bibnamefont{Sen}},
  \bibinfo{journal}{Phys. Rev. B} \textbf{\bibinfo{volume}{77}},
  \bibinfo{pages}{214427} (\bibinfo{year}{2008}).

\bibitem[{\citenamefont{Roux}(2009)}]{roux_time_dependent_lanczos}
\bibinfo{author}{\bibfnamefont{G.}~\bibnamefont{Roux}}, \bibinfo{journal}{Phys.
  Rev. A} \textbf{\bibinfo{volume}{79}}, \bibinfo{pages}{021608(R)}
  (\bibinfo{year}{2009}).

\bibitem[{qua(2008)}]{quantum_engineering}
\bibinfo{journal}{See, for example, Science} \textbf{\bibinfo{volume}{320}},
  \bibinfo{pages}{312} (\bibinfo{year}{2008}).

\bibitem[{\citenamefont{Greiner
  et~al.}(2002{\natexlab{a}})\citenamefont{Greiner, Mandel, Esslinger,
  Ha\"ensch, and Bloch}}]{greiner_transition_superfluid_mott}
\bibinfo{author}{\bibfnamefont{M.}~\bibnamefont{Greiner}},
  \bibinfo{author}{\bibfnamefont{O.}~\bibnamefont{Mandel}},
  \bibinfo{author}{\bibfnamefont{T.}~\bibnamefont{Esslinger}},
  \bibinfo{author}{\bibfnamefont{T.~W.} \bibnamefont{Ha\"ensch}},
  \bibnamefont{and} \bibinfo{author}{\bibfnamefont{I.}~\bibnamefont{Bloch}},
  \bibinfo{journal}{Nature (London)} \textbf{\bibinfo{volume}{415}},
  \bibinfo{pages}{39} (\bibinfo{year}{2002}{\natexlab{a}}).

\bibitem[{\citenamefont{Greiner
  et~al.}(2002{\natexlab{b}})\citenamefont{Greiner, Mandel, H\"ansch, , and
  Bloch}}]{greiner_fast_tunnability}
\bibinfo{author}{\bibfnamefont{M.}~\bibnamefont{Greiner}},
  \bibinfo{author}{\bibfnamefont{O.}~\bibnamefont{Mandel}},
  \bibinfo{author}{\bibfnamefont{T.}~\bibnamefont{H\"ansch}}, ,
  \bibnamefont{and} \bibinfo{author}{\bibfnamefont{I.}~\bibnamefont{Bloch}},
  \bibinfo{journal}{Nature (London)} \textbf{\bibinfo{volume}{419}},
  \bibinfo{pages}{51} (\bibinfo{year}{2002}{\natexlab{b}}).

\bibitem[{\citenamefont{Kinoshita et~al.}(2006)\citenamefont{Kinoshita, Wenger,
  and Weiss}}]{kinoshita_non_thermalization}
\bibinfo{author}{\bibfnamefont{T.}~\bibnamefont{Kinoshita}},
  \bibinfo{author}{\bibfnamefont{T.}~\bibnamefont{Wenger}}, \bibnamefont{and}
  \bibinfo{author}{\bibfnamefont{D.~S.} \bibnamefont{Weiss}},
  \bibinfo{journal}{Nature (London)} \textbf{\bibinfo{volume}{440}},
  \bibinfo{pages}{900} (\bibinfo{year}{2006}).

\bibitem[{\citenamefont{Lieb and Liniger}(1963)}]{lieb_liniger_model}
\bibinfo{author}{\bibfnamefont{E.~H.} \bibnamefont{Lieb}} \bibnamefont{and}
  \bibinfo{author}{\bibfnamefont{W.}~\bibnamefont{Liniger}},
  \bibinfo{journal}{Phys. Rev.} \textbf{\bibinfo{volume}{130}},
  \bibinfo{pages}{1605} (\bibinfo{year}{1963}).

\bibitem[{\citenamefont{Jaynes}(1957{\natexlab{a}})}]{JaynesI}
\bibinfo{author}{\bibfnamefont{E.~T.} \bibnamefont{Jaynes}},
  \bibinfo{journal}{Phys. Rev.} \textbf{\bibinfo{volume}{106}},
  \bibinfo{pages}{620} (\bibinfo{year}{1957}{\natexlab{a}}).

\bibitem[{\citenamefont{Jaynes}(1957{\natexlab{b}})}]{JaynesII}
\bibinfo{author}{\bibfnamefont{E.~T.} \bibnamefont{Jaynes}},
  \bibinfo{journal}{Phys. Rev.} \textbf{\bibinfo{volume}{108}},
  \bibinfo{pages}{171} (\bibinfo{year}{1957}{\natexlab{b}}).

\bibitem[{\citenamefont{Srednicki}(1994)}]{Srednicki_eigenstate}
\bibinfo{author}{\bibfnamefont{M.}~\bibnamefont{Srednicki}},
  \bibinfo{journal}{Phys. Rev. E} \textbf{\bibinfo{volume}{50}},
  \bibinfo{pages}{888} (\bibinfo{year}{1994}).

\bibitem[{\citenamefont{Iucci and Cazalilla}(2009)}]{unpub}
\bibinfo{author}{\bibfnamefont{A.}~\bibnamefont{Iucci}} \bibnamefont{and}
  \bibinfo{author}{\bibfnamefont{M.~A.} \bibnamefont{Cazalilla}},
  \bibinfo{journal}{{\it in preparation}}  (\bibinfo{year}{2009}).

\bibitem[{\citenamefont{Barthel and Schollw\"ock}(2008)}]{schollwoeck_quench}
\bibinfo{author}{\bibfnamefont{T.}~\bibnamefont{Barthel}} \bibnamefont{and}
  \bibinfo{author}{\bibfnamefont{U.}~\bibnamefont{Schollw\"ock}},
  \bibinfo{journal}{Phys. Rev. Lett.} \textbf{\bibinfo{volume}{100}},
  \bibinfo{pages}{100601} (\bibinfo{year}{2008}).

\bibitem[{\citenamefont{Luttinger}(1963)}]{luttinger_model}
\bibinfo{author}{\bibfnamefont{J.~M.} \bibnamefont{Luttinger}},
  \bibinfo{journal}{J. Math. Phys.} \textbf{\bibinfo{volume}{4}},
  \bibinfo{pages}{1154} (\bibinfo{year}{1963}).

\bibitem[{\citenamefont{Mattis and Lieb}(1965)}]{mattis_lieb_luttinger_model}
\bibinfo{author}{\bibfnamefont{D.~C.} \bibnamefont{Mattis}} \bibnamefont{and}
  \bibinfo{author}{\bibfnamefont{E.~H.} \bibnamefont{Lieb}},
  \bibinfo{journal}{J. Math. Phys.} \textbf{\bibinfo{volume}{6}},
  \bibinfo{pages}{304} (\bibinfo{year}{1965}).

\bibitem[{\citenamefont{Tomonaga}(1950)}]{Tomonaga_1D_electron_gas}
\bibinfo{author}{\bibfnamefont{S.}~\bibnamefont{Tomonaga}},
  \bibinfo{journal}{Prog. Theor. Phys.} \textbf{\bibinfo{volume}{5}},
  \bibinfo{pages}{544} (\bibinfo{year}{1950}).

\bibitem[{\citenamefont{Luther and
  Peschel}(1974)}]{luther_peschel_correlation_functions}
\bibinfo{author}{\bibfnamefont{A.}~\bibnamefont{Luther}} \bibnamefont{and}
  \bibinfo{author}{\bibfnamefont{I.}~\bibnamefont{Peschel}},
  \bibinfo{journal}{Phys. Rev. B} \textbf{\bibinfo{volume}{9}},
  \bibinfo{pages}{2911} (\bibinfo{year}{1974}).

\bibitem[{\citenamefont{Haldane}(1980)}]{haldane_exponents_spin_chain}
\bibinfo{author}{\bibfnamefont{F.~D.~M.} \bibnamefont{Haldane}},
  \bibinfo{journal}{Phys. Rev. Lett.} \textbf{\bibinfo{volume}{45}},
  \bibinfo{pages}{1358} (\bibinfo{year}{1980}).

\bibitem[{\citenamefont{Haldane}(1981{\natexlab{a}})}]{haldane_luttinger_liqui%
d}
\bibinfo{author}{\bibfnamefont{F.~D.~M.} \bibnamefont{Haldane}},
  \bibinfo{journal}{J. Phys. C: Solid State Phys.}
  \textbf{\bibinfo{volume}{14}}, \bibinfo{pages}{2585}
  (\bibinfo{year}{1981}{\natexlab{a}}).

\bibitem[{\citenamefont{Haldane}(1981{\natexlab{b}})}]{haldane_effective_harmo%
nic_fluid_approach}
\bibinfo{author}{\bibfnamefont{F.~D.~M.} \bibnamefont{Haldane}},
  \bibinfo{journal}{Phys. Rev. Lett.} \textbf{\bibinfo{volume}{47}},
  \bibinfo{pages}{1840} (\bibinfo{year}{1981}{\natexlab{b}}).

\bibitem[{\citenamefont{Giamarchi}(2004)}]{giamarchi_book_1d}
\bibinfo{author}{\bibfnamefont{T.}~\bibnamefont{Giamarchi}},
  \emph{\bibinfo{title}{Quantum physics in one dimension}}
  (\bibinfo{publisher}{Oxford University Press}, \bibinfo{address}{Oxford},
  \bibinfo{year}{2004}).

\bibitem[{\citenamefont{Gradshteyn and Ryzhik}(1980)}]{gradshteyn80_tables}
\bibinfo{author}{\bibfnamefont{A.}~\bibnamefont{Gradshteyn}} \bibnamefont{and}
  \bibinfo{author}{\bibfnamefont{R.}~\bibnamefont{Ryzhik}},
  \emph{\bibinfo{title}{Tables of integrals series and products}}
  (\bibinfo{publisher}{Academic Press}, \bibinfo{address}{New-York},
  \bibinfo{year}{1980}).

\bibitem[{\citenamefont{Balian}(1991)}]{balian_stat_mech}
\bibinfo{author}{\bibfnamefont{R.}~\bibnamefont{Balian}},
  \emph{\bibinfo{title}{From Microphysics to Macrophysics: Applications of
  Statistical Mechanics, vol. I}} (\bibinfo{publisher}{Springer-Verlag},
  \bibinfo{address}{Berlin}, \bibinfo{year}{1991}).

\bibitem[{\citenamefont{G\"unter et~al.}(2005)\citenamefont{G\"unter,
  St\"oferle, Moritz, K\"ohl, and
  Esslinger}}]{gunter_p_wave_interactions_1D_fermions}
\bibinfo{author}{\bibfnamefont{K.}~\bibnamefont{G\"unter}},
  \bibinfo{author}{\bibfnamefont{T.}~\bibnamefont{St\"oferle}},
  \bibinfo{author}{\bibfnamefont{H.}~\bibnamefont{Moritz}},
  \bibinfo{author}{\bibfnamefont{M.}~\bibnamefont{K\"ohl}}, \bibnamefont{and}
  \bibinfo{author}{\bibfnamefont{T.}~\bibnamefont{Esslinger}},
  \bibinfo{journal}{Phys. Rev. Lett.} \textbf{\bibinfo{volume}{95}},
  \bibinfo{pages}{230401} (\bibinfo{year}{2005}).

\bibitem[{\citenamefont{St{\"o}ferle et~al.}(2004)\citenamefont{St{\"o}ferle,
  Moritz, Schori, K{\"o}hl, and
  Esslinger}}]{stoeferle_shaking_fast_tunnability}
\bibinfo{author}{\bibfnamefont{T.}~\bibnamefont{St{\"o}ferle}},
  \bibinfo{author}{\bibfnamefont{H.}~\bibnamefont{Moritz}},
  \bibinfo{author}{\bibfnamefont{C.}~\bibnamefont{Schori}},
  \bibinfo{author}{\bibfnamefont{M.}~\bibnamefont{K{\"o}hl}}, \bibnamefont{and}
  \bibinfo{author}{\bibfnamefont{T.}~\bibnamefont{Esslinger}},
  \bibinfo{journal}{Phys. Rev. Lett.} \textbf{\bibinfo{volume}{92}},
  \bibinfo{pages}{130403} (\bibinfo{year}{2004}).

\bibitem[{\citenamefont{Mathey et~al.}(2008)\citenamefont{Mathey, Altman, and
  Vishwanath}}]{mathey_noise_correlations}
\bibinfo{author}{\bibfnamefont{L.}~\bibnamefont{Mathey}},
  \bibinfo{author}{\bibfnamefont{E.}~\bibnamefont{Altman}}, \bibnamefont{and}
  \bibinfo{author}{\bibfnamefont{A.}~\bibnamefont{Vishwanath}},
  \bibinfo{journal}{Phys. Rev. Lett.} \textbf{\bibinfo{volume}{100}},
  \bibinfo{pages}{240401} (\bibinfo{year}{2008}).

\bibitem[{\citenamefont{Polkovnikov et~al.}(2006)\citenamefont{Polkovnikov,
  Altman, and Demler}}]{polkovnikov_interference_between_condensates}
\bibinfo{author}{\bibfnamefont{A.}~\bibnamefont{Polkovnikov}},
  \bibinfo{author}{\bibfnamefont{E.}~\bibnamefont{Altman}}, \bibnamefont{and}
  \bibinfo{author}{\bibfnamefont{E.}~\bibnamefont{Demler}},
  \bibinfo{journal}{Proc. Natl. Acad. Sci. USA} \textbf{\bibinfo{volume}{103}},
  \bibinfo{pages}{6125} (\bibinfo{year}{2006}).

\bibitem[{\citenamefont{Ma}(1985)}]{ma_stat_mech}
\bibinfo{author}{\bibfnamefont{S.-K.} \bibnamefont{Ma}},
  \emph{\bibinfo{title}{Statistical Mechanics}} (\bibinfo{publisher}{World
  Scientific}, \bibinfo{address}{Singapore}, \bibinfo{year}{1985}).

\end{thebibliography}

\end{document}